\documentclass{aa}  
\usepackage{graphicx}
\usepackage{txfonts}
\usepackage{natbib}
\usepackage{longtable}
\bibpunct{(}{)}{;}{a}{}{,}   

\begin{document}
   \title{Spectra probing the number ratio of C- to M-type AGB stars in the NGC 6822 galaxy
\thanks{Based on observations made with ESO telescopes at the La Silla Paranal Observatory under programme ID 383.D-0367}$^,$
\thanks{Complete Table 2 and the reduced spectra in ASCII format are available at the CDS via anonymous ftp to cdsarc.u-strasbg.fr (130.79.128.5).}
}

%
   
\author{N.~Kacharov\inst{1,2,3}
	\and M.~Rejkuba\inst{1}
	\and M.-R.~L.~Cioni\inst{4,5}\thanks{Research Fellow of the Alexander von Humboldt Foundation}
}

   \offprints{N. Kacharov, kacni@abv.bg}

   \institute{ESO, Karl-Schwarzschild-Strasse 2, D-85748 Garching, Germany
          \and
	      Institute of Astronomy, Bulgarian Academy of Sciences, 72 Tsarigradsko Chaussee Blvd., 1784 Sofia, Bulgaria
          \and
	      Department of Astronomy, St. Kliment Ohridski University of Sofia, 5 James Bourchier Blvd., 1164 Sofia, Bulgaria
	  \and 
              University Observatory Munich, Scheinerstrasse 1, 81679 M\"{u}nchen, Germany
          \and 
              University of Hertfordshire, Physics Astronomy and Mathematics, Hatfield AL10 9AB, United Kingdom
             }

   \date{Received: 31.05.2011 / Accepted: 16.10.2011}
\titlerunning{AGB stellar population in NGC 6822}

 
  \abstract
   {}
   {We calibrate spectroscopically the C- versus (vs.) M-type asymptotic giant branch (AGB) star selection made using near-IR photometry, and investigate the spatial distribution of the C/M ratio in NGC 6822, based on low resolution spectroscopy and near-IR photometry.}
   {We obtained low resolution multi-object spectroscopy with the VIMOS instrument at the ESO VLT of $\sim800$ stars in seven fields centred on NGC~6822. The spectroscopic classification of giant stars in NGC 6822 and foreground dwarf contaminants was  made by comparing more than 500 good quality spectra with the spectroscopic atlas of Turnshek et al. (1985). The sample of spectroscopically confirmed AGB stars in NGC 6822 is divided into C- and M-rich giants to constrain the C vs. M AGB star selection criteria based on photometry. The larger near-IR photometric sample is then used to investigate the C/M ratio gradients across the galaxy.}
   {We present the largest catalogue of near-IR photometry and spectra of AGB stars in NGC 6822 with 150 C-stars and 122 M-stars.  
Seventy-nine percent of the C-stars in our catalogue are redder than $(J-K)_0=1.2$~mag, and $12\%$ are brighter than $K_0=16.45$~mag and bluer than $(J-K)_0=1.2$~mag. The remaining $9\%$ are mixed with the M-type AGB stars, $88\%$ of which have colours $(J-H)_0>0.73$~mag and $(J-K)_0$ between $0.9$~mag and $1.2$~mag. The remainder are mixed with dwarfs and C-type stars. The foreground dwarfs have preferably colours $(J-H)_0<0.73$~mag~$(95\%)$. Using the proposed criteria, we estimate that the overall C/M ratio of the galaxy is around $0.8$ with a spread between $0.2<$~C/M~$<1.8$. These results suggest that the metallicity index [Fe/H] is between $-1.2$~dex and $-1.3$~dex according to the different calibrations and that there is a significant spread of about $0.4\div0.6$~dex. We also discuss age rather than metallicity variations that could explain the C/M ratio trends.} 
  {}


   \maketitle
%

\section{Introduction}

The intermediate-age asymptotic giant branch (AGB) stars are classified as carbon-rich (C stars) or oxygen-rich (M stars), depending on which element dominates their atmosphere.  These stars are among the most luminous stars in a galaxy in near-IR, and are thus easily observable at large distances. The upcoming large telescopes ($>20$~m) will operate most efficiently in near-IR and will detect large samples of AGB stars in nearby galaxies. In Addition, this stellar evolutionary phase is extremely important to the studies of high redshift galaxies, because these relatively young galaxies contain significant fraction of intermediate-age stars, and in a 1~Gyr old simple stellar population up to 80\% of the K-band light originates from the luminous AGB stars \citep{maraston05, maraston+06}. Nevertheless, the intermediate-age stellar evolutionary models, the stellar atmospheres, the internal composition, and the physics of luminous AGB stars are all affected by large uncertainties \citep{cassisi+01,gallart+05,ventura+marigo10}, and currently this is one area of stellar evolutionary theory where important improvements are needed, both theoretically \citep{marigo+girardi07,marigo+08} and observationally \citep[e.g.][]{girardi+marigo07,lyubenova+10,girardi+10}.


At present, the selection of C and M stars is made using either near-IR colour-magnitude diagrams or colour-colour diagrams for broad- and narrow-band filters \citep{battinelli2004}. These surveys rely on photometric selection criteria that have inherent uncertainties, do not easily distinguish K giants and Galactic M dwarfs, and do not account for S-type stars. The situation somewhat improves, at least for K giants, if optical and IR selection criteria are used together \citep{cioni+habing05}. The spectroscopic classification of AGB stars covering a range of metallicities is necessary to properly estimate the biases and establish quantitative criteria for the photometric selection boxes.

The C/M ratio is the number ratio of C-type (carbon-rich) to M-type (oxygen-rich) AGB stars. It is a function of the [Fe/H] abundance \citep{battinelli2005} and provides a simple indication of the metallicity distribution across galaxies \citep[e.g.][for M 33 and the Magellanic Clouds, respectively]{cioni2008,cioni2009,cioni+habing03}. 
The C/M vs. [Fe/H] relation is due to the combined effect of pronounced carbon 
dredge-up on the stellar spectrum at low metallicity, and the blue colours of metal-poor red giants \citep{iben1983}. The lower metallicity O-rich stars turn into C-type AGB stars more easily and remain so for a longer time than those of higher metallicity because fewer carbon atoms need to be dredged up to effect this transformation, and in addition at lower metallicity the AGB evolutionary tracks have higher temperatures, which causes lower abundances of TiO molecules, hence smaller numbers of M stars. However, while theoretically clear arguments exist for the dependence of C/M ratio on [Fe/H],  the relation is still poorly calibrated owing to: (i) the ill-defined criteria for selecting C and M stars; (ii) the contamination from both foreground and S stars (S stars have equal amounts of carbon and oxygen in their atmospheres); and (iii) the need to have a well-defined sample of AGB stars with spectroscopically determined metallicites. 

The NGC 6822 galaxy is an ideal target for the study of C- and M-type AGB stars and the C/M ratio as a function of metallicity. 
It is a relatively well-studied dwarf irregular galaxy in the constellation Sagittarius at a distance of $\sim490$~kpc \citep{mateo98}. Owing to its low Galactic latitude ($b = -18.39^\circ$), it is affected by moderate foreground extinction, $E(B-V)=0.24$~mag \citep{schlegel+98}. It is similar in mass ($1.9\times10^9~M_{\odot}$), structure, and metallicity ($\sim 0.2~Z_{\odot}$) to the Small Magellanic Cloud and consists of different morphological components (disk, bar, and halo), rotation, gas content, active star formation, and dark matter distribution \citep{weldrake2003}. Star formation in this galaxy began at least $10$~Gyr ago \citep{clementini2003} and remains ongoing \citep{gallart+96, hutchings1999}. The galaxy is embedded in a large HI envelope \citep{weldrake2003}. Previous spectroscopic studies, although incomplete, of supergiants \citep{venn2001}, HII regions \citep{chandar2000}, and red giant branch (RGB) stars \citep{tolstoy2001} found evidence of a gradient in [O/H] and measured an average [Fe/H] $= -0.9$~dex with a spread of $1.5$~dex.
Most importantly, this galaxy has a large and widely distributed intermediate-age population \citep{battinelli2006} with a clear spread in metallicity \citep{cioni+habing05}. The C/M ratio has been estimated photometrically both from optical and near-IR criteria.
This work provides the first comprehensive {\it spectroscopic} study of the AGB population of NGC 6822 dwarf irregular galaxy and its metallicity distribution. 

We initiated a project to improve the C/M vs. [Fe/H] calibration using the large sample of intermediate-age stars in NGC 6822, for which we have a large database of near-IR photometry, low-resolution optical spectra useful to spectroscopically select C- and M-type AGB stars, as well as near-IR Ca II triplet spectroscopic observations used to indirectly measure [Fe/H]. Our preliminary results for the near-IR photometry were presented by \citet{sibbons+10}, and their full analysis is in preparation. 
This work analyses the near-IR optical spectra obtained with VIMOS multi-object spectrograph at the ESO VLT, and in a future article we will present the near-IR Ca II spectroscopic observations of AGB stars in NGC 6822.

\section{Observations and data reduction}

Our spectroscopic survey of M- and C-type AGB stars in the dwarf irregular galaxy NGC 6822 was made using the VIMOS imager and multi-object spectrograph \citep{lefevre03} at the ESO VLT UT3 telescope. The instrument is mounted on the Nasmyth B focus of UT3 Melipal and has four identical arms, each with a $\sim7' \times 8'$ FOV and a $0\farcs205$ pixel size. The gap between the quadrants is $\sim 2'$ and each quadrant is equipped with one EEV $2\mathrm{k} \times 4\mathrm{k}$ CCD.

We first obtained mandatory R-band pre-imaging observations of seven fields in NGC~6822 in service mode. These pre-images were then used together with the wide field near-IR photometric catalogue obtained at UKIRT with WFCAM \citep[][Sibbons et al. 2011, in prep]{sibbons+10}, to prepare the multi-object masks for spectroscopic follow-up observations in visitor mode, which comprised four masks (one per VIMOS quadrant) for each pointing.

\begin{table*}
\begin{center}
\caption{Observing log.}\label{tab:Obs_log}
{\normalsize
 \begin{tabular}{ccccccc}
\hline
Night & Field & RA  & DEC  & Exposure & Slits & Stars \\
      &       & [hh:mm:ss] & [$\,^{\circ}:\arcmin:\arcsec$] & [s] & & \\
\hline
 22-23 Aug. 2009 & CF1 & 19:44:56.75 & $-$14:48:14.18 & $2 \times 1200$ & 119 & 57 \\
                 & CF2 & 19:44:56.75 & $-$14:48:14.18 & $2 \times 1200$ & 108 & 81 \\
                 & NW1 & 19:44:20.77 & $-$14:40:53.26 & $2 \times 1200$ & 108 & 47 \\
                 & NE1 & 19:45:31.78 & $-$14:40:52.64 & $2 \times 1200$ & 114 & 96 \\
                 & SE1 & 19:45:33.78 & $-$14:55:45.77 & $2 \times 1200$ & 123 &115 \\
 23-24 Aug. 2009 & SW1 & 19:44:19.78 & $-$14:55:43.03 & $2 \times 1200$ & 101 & 62 \\
                 & NE2 & 19:48:17.18 & $-$14:24:45.94 & $2 \times 1200$ &  78 & 70 \\
                 & SW2 & 19:42:09.78 & $-$15:14:40.20 & $2 \times 1200$ &  70 & 18 \\
\hline
 \end{tabular}
\par}
\end{center}
\end{table*}

We targeted seven fields, four centred on NGC~6822, and two outer fields. The central field had two different mask sets, each with 100-120 targets, while the other six fields had each one set of masks. In total, we targeted $\sim 800$ stars in eight setups. The spectroscopic target selection was based on the UKIRT near-IR photometric data.  All discussed magnitudes and colours in this study were corrected for foreground extinction according to the \citet{schlegel+98} extinction map.  All selected targets are brighter than $17.45$~mag in $K_0$ and have $(J-K)_0 > 0.74$~mag, although redder stars were targeted preferentially. \citet{cioni+habing05} detected the RGB tip (TRGB) at $K_s=17.10\pm0.01$~mag, while \citet{sibbons+10} report that it varies across the galaxy by $\Delta K = 1.36$~mag, with an average value of $K = 17.48\pm0.26$~mag. \citet{davidge2003} adopted a distance modulus of $23.49$~mag based on the Cepheid and RGB tip measurements of \citet{gallart+96b}, and  measured the onset of the RGB tip in $K$ band near $K=17$ mag. This is consistent with $M_K^{\mathrm{RGBT}} \sim -6.5$~mag. \citet{gorski2011} identified the TRGB at $K = 16.97\pm0.09$~mag.

Our observing log is presented in Table \ref{tab:Obs_log}. In the second last column (slits), we report the number of slits for each pointing (sum of slits in four VIMOS quadrants), and in the last column (stars) we list the number of good quality recorded spectra (quality flags: 4, 5, or 6; see below) for each observed setup.

For our spectroscopic observations on August 22-24, 2009, we used the medium resolution grism (MR grism)  and GG475 order sorting filter, which provides a spectral resolution $R=580$ ($2.5$ \AA/pix dispersion) and a wavelength coverage from $500$ to $1000$~nm.  The severe fringing in the red for the old VIMOS thinned and back-side illuminated, single-layer coated CCDs, used until May 2010 \citep{hammersley+10}, meant that there was a significantly more restricted useful wavelength range of 500-780 nm in our spectra. Each setup was exposed for $2 \times 20$~min.


We used the ESO VIMOS pipeline (version 2.5.2) to reduce the spectra. For each of the two observing nights, we took separate sets of bias frames and spectrophotometric standard-star observations. The reduction and extraction of the scientific exposures was performed in three steps with the following pipeline recipes. First we used the $vmbias$ recipe to create a master bias frame averaging the five bias frames. The $vmmoscalib$ recipe was then used to obtain the wavelength calibration from the arc lamp spectra, trace the edges of the associated flat fields, and prepare all necessary tables for the scientific extraction.

We note that the flat-field exposures were only used to trace the slit edges and prepare the extraction tables. We decided not to apply a flat-field correction to our target spectra, as the flat fields were taken in the morning after the observing night. The different rotator angle and small differences in the positioning of the masks in the focal plane of the instrument may cause small shifts between the flat-field and science exposures, which could then cause a decrease in the signal-to-noise ratio (S/N) after flat-fielding. We experimented with applying the flat-field correction but found that this neither improved nor decreased the spectrum S/N in the blue, while in the red part owing to fringing, spectra that were flat-fielded were actually noisier. We only find a difference in the third quadrant, where all exposures display some artificial drop in efficiency for certain wavelength ranges (most significantly around the sodium doublet, as most clearly seen in Fig. \ref{fig:M0_Dwarfs}), which could not be fully corrected by applying a flat-field correction. This, however, did not affect our spectral classification.

\begin{figure}
\centering
\resizebox{\hsize}{!}{
\includegraphics[angle=0]{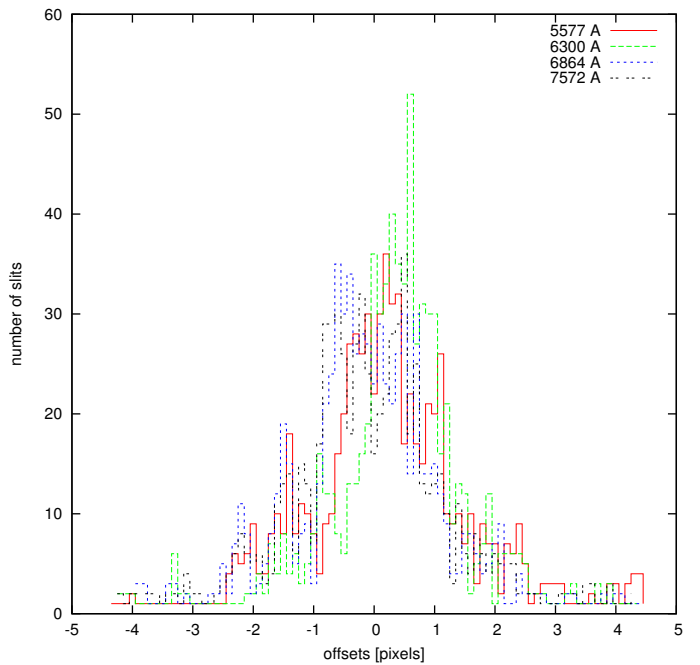}
}
\caption[]{Distribution of offsets in pixels for four sky lines with respect to their expected wavelengths.
}\label{fig:all_offsets}
\end{figure}

The $vmmosscience$ recipe was used to extract and calibrate the target spectra. This recipe applies the extraction mask obtained with $vmmoscalib$. The slit spectra are bias subtracted and remapped by eliminating the optical distortions. An additional wavelength calibration adjustment was made by fitting several strong sky emission lines. The final mean model accuracy of the wavelength calibration is about $0.15$~pix, but the measurements of sky line offsets with respect to the expected wavelength sometimes have values of up to several pixels for individual spectra (Fig.~\ref{fig:all_offsets}). 

We had acquired two scientific exposures for each observed field, which were aligned and stacked together. We adopted local sky subtraction and cosmic cleaning, and the optimal extraction method \citep{horne86} within the $vmmosscience$ recipe. Finally, we applied the response curve, obtained from a spectrophotometric standard star observation. The reduction process of the standard star observations was analogous to the scientific ones. During the entire reduction process, we manually checked all spectra to verify spectral tracing, wavelength calibration, and object detection.



Unfortunately, the low resolution of our spectra and the very broad spectral features, prevented accurate measurements of the radial velocities of the targeted stars. On the basis of the Besan\c{c}on model of our Galaxy \citep{robin+03}, the expected median radial velocity and its standard deviation for the Milky Way foreground population of stars with $15 < K < 17.5$~mag in the direction of NGC~6822 is $1 \pm 44$~km/s. Owing to the overlap of this distribution with the radial velocity of NGC 6822 ($-57$~km/s), which is based on HI data \citep{koribalski+04}, and the radial velocities of the carbon stars obtained by \citet{demers2006} (between $+10$ and $-70$~km/s $\pm15$~km/s), combined with the uncertainty in our measured radial velocities, we decided not to rely on radial velocities to distinguish between the foreground dwarfs and NGC 6822 giant star members, but to instead use spectral features to distinguish the two populations. The classification of spectral types for all acquired spectra is described in Sect. 3.

\begin{figure}
\centering
\resizebox{\hsize}{!}{
\includegraphics[angle=0]{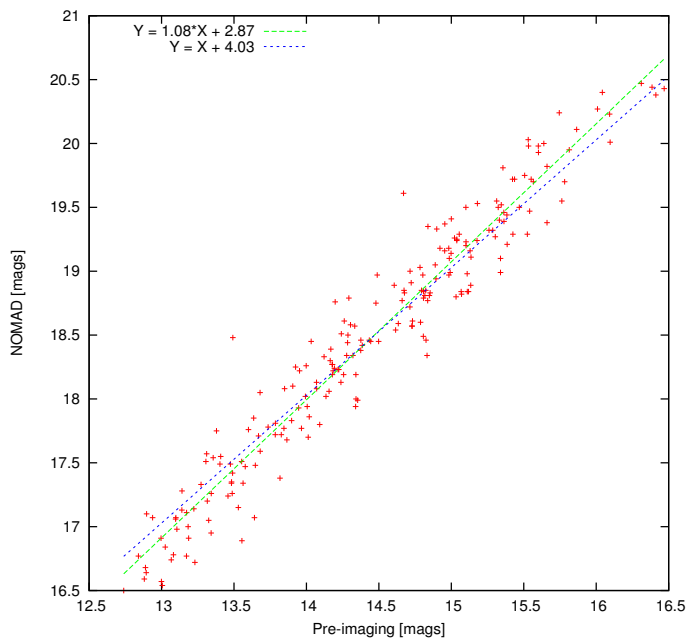}
}
\caption[]{R-band pre-imaging calibration. The lines are fits to the data indicated by the equations given within the figure.
}
\label{fig:calib}
\end{figure}

We used SExtractor \citep{bertin+96} to derive R-band photometry from the pre-imaging, which was then calibrated using the NOMAD catalogue \citep{zacharias+2005} based on about 200 stars in common with our observations. The calibration is presented in Fig. \ref{fig:calib}. We tested two fits to the data, a linear and a constant shift, which are presented in Eq. 1 and 2.

\begin{equation}
 R_{\mathrm{NOMAD}} = 1.08^{\pm0.02}\times R_{\mathrm{pre-imaging}} + 2.87^{\pm0.27} ~ (RMS = 0.24),
\end{equation}

\begin{equation}
 R_{\mathrm{NOMAD}} = R_{\mathrm{pre-imaging}} + 4.03^{\pm0.02} ~~~~~~~~~~~~~~~~~~~ (RMS = 0.25).
\end{equation}

In obtaining the R-band photometry, we used the linear fit calibration (Eq. 1). All R-magnitudes were corrected for interstellar extinction according to \citet{schlegel+98} using the NED extinction-law calculator. We note that the errors in the R-band photometry are larger than the errors of the IR-photometry because the NOMAD catalogue is a compilation of data from different sources.

\section{Spectral classification}

For each of our spectroscopic targets, we compiled near-IR photometric data based on the UKIRT photometry \citep{sibbons+10} and complemented this with the spectral type based on reduced VIMOS spectra. 

\begin{figure}
\centering
\resizebox{\hsize}{!}{
\includegraphics[angle=0]{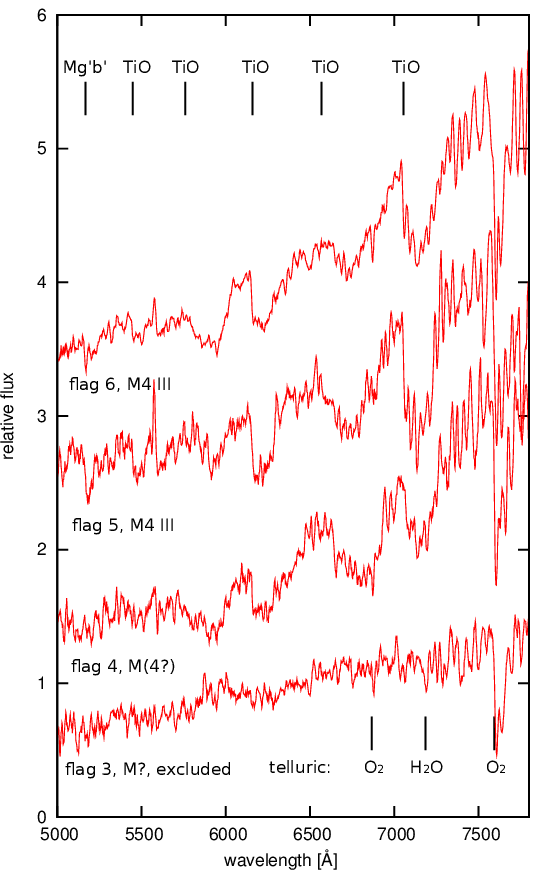}
}
\caption[]{Example of individual spectra with different quality flags. The spectra are rescaled, shifted along the y-axis by an arbitrary constant for presentation purpose, and smoothed with a level 5 boxcar function.
}
\label{fig:quality}
\end{figure}

We determined the spectral class of each star mainly by visual comparing our spectra with those in the ``Atlas of digital spectra of cool stars`` \citep{atlas}. Where possible, we also compared our spectra with the spectral library of \citet{jakoby84}. This spectral library is unfortunately less complete, in particular, there are not C-rich giants. To match the resolution of our spectra to that of the spectral atlas, and reduce the noise, we smoothed our spectra with a level 5 boxcar smoothing function using the {\it splot} IRAF task. Each spectrum was assigned one of the following quality flags, which are based primarily on the clarity of the spectral features, and not on the S/N: flag 2 for no detection; flag 3 for a poor quality spectrum, the star cannot be classified; flag 4, when we can determine whether the star is M- or C-type and eventually whether it is a giant or a foreground dwarf; flag 5 for very good quality spectra, the error in the spectral classification is within $\pm$ one subtype; and flag 6 for an excellent quality and precise determination of the spectral class. The stars with lowest quality flags (2 or 3) were excluded from our final list. Fig.~\ref{fig:quality} shows typical examples of spectra with different quality flags. We indicate in this figure on top the primary spectral features that were used to assign a spectral type to M giants. 

Typical M giants are characterized by TiO absorption bands, which are stronger for increasingly later spectral types. The telluric features $\mathrm{O}_2$ and $\mathrm{H}_2\mathrm{O}$ are present in all spectra, because we did not apply a telluric correction, and are indicated at the bottom of the figure. The C-type giants can be quite easily distinguished as they have characteristic $\mathrm{C}_2$ and CN molecular bands. The S-type stars display ZrO absorption features. 

\begin{figure}
\centering
\resizebox{\hsize}{!}{
\includegraphics[angle=0]{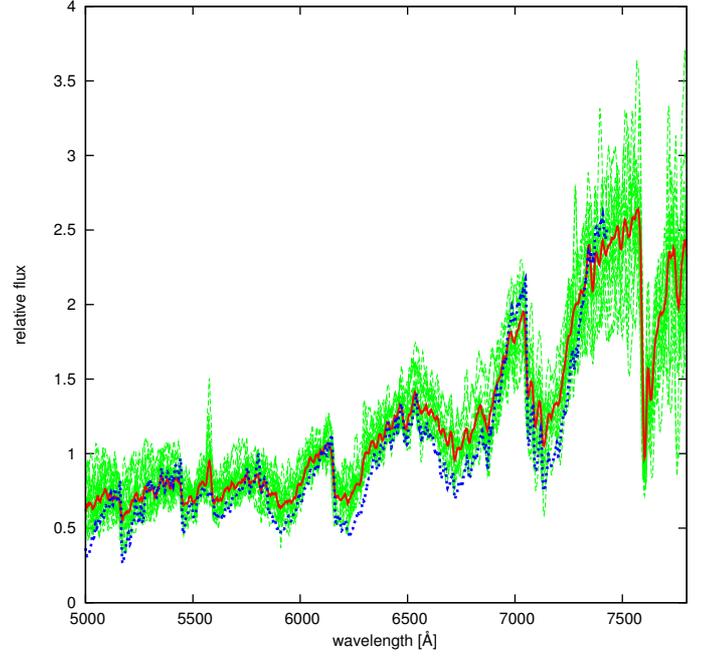}
}
\caption[]{All spectra classified as M4 III are rescaled and plotted together with green lines. An average M4 III spectrum is calculated and shown in red and a reference M4 III spectrum from the library of \citet{jakoby84} is drawn with a blue dotted line.
}
\label{fig:M4_Giants}
\end{figure}

\begin{figure}
\centering
\resizebox{\hsize}{!}{
\includegraphics[angle=0]{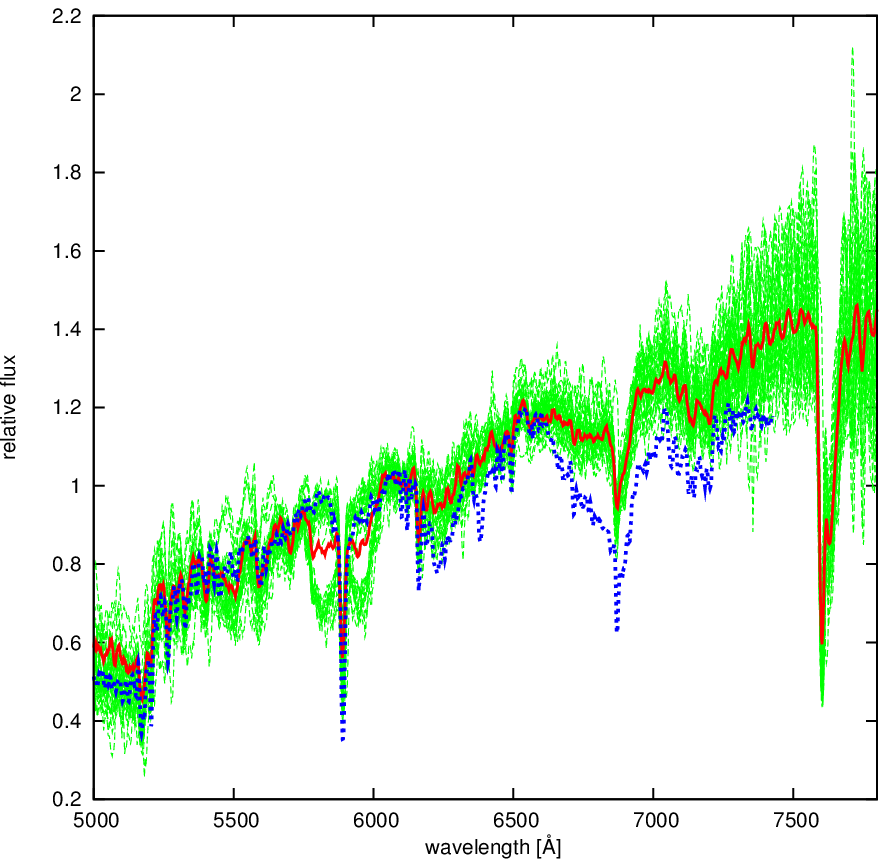}
}
\caption[]{Same as Fig. \ref{fig:M4_Giants} but for M0 V spectral class.
}
\label{fig:M0_Dwarfs}
\end{figure}

After we had performed an initial ``by eye'' spectral classification, we rescaled all spectra to an equal flux level and plotted them separately for each spectral sub-class to verify that indeed all spectra that had been assigned to that spectral sub-class had similar spectra. If an outlier was found, its spectral classification was repeated by comparing its spectrum with average spectra for different adjacent spectral classes. For each spectral subtype, we compiled an average spectrum. Fig.~\ref{fig:M4_Giants} compares spectra of all M4 giants (green lines) with the spectrum from \citet{jakoby84} (dotted blue line), as well as the average M4 giant spectrum (red thick line). Similarly, Fig.~\ref{fig:M0_Dwarfs} compares M0 dwarf spectra (green) with the average spectrum (red) and the M0 dwarf spectrum from \citet{jakoby84} (dotted blue line). We note that the molecular bands of the reference spectra from \citet{jakoby84} library are deeper, than those of average NGC 6822 spectra, most probably because of the higher metallicity of Milky Way stars. We tried to find information about the metallicities of the used reference stars in the Pastel database \citep{pastel} but found data for only two reference giants, which both have solar metallicity. A comparison of synthetic spectra of different metallicities from the standard stellar library of \citet{lejeune+97} reveals some differences in the depths of the main molecular bands and the shapes of the spectra. The stars of higher metallicity appear to have slightly later subtype than those of lower metallicity, at a given temperature and gravity. This means that our spectral classification of the M-type giants might be slightly underestimated, because of the lower average metallicity of NGC 6822 with respect to stars in the comparison spectral atlases \citep{atlas,jakoby84}, which are principally located in the solar neighbourhood.

The wide absorption feature around the Na I  line in some of the M dwarf spectra (shown with green lines) in Fig. \ref{fig:M0_Dwarfs} is caused by flat field residuals. All spectra displaying these artifacts were acquired on the third quadrant CCD. We note that they are not real features. We note that all bands in the \citet{jakoby84} M0 V spectrum appear deeper than in our M0 V spectra. This is due to differences between the reference libraries. The M0 V spectral class of \citet{jakoby84} is indeed more consistent with the M1 V spectral class of \citet{atlas}, which is our primary reference. Unfortunately, we could not find information about the metallicities of these reference stars in the literature.

Table \ref{tab:classification} shows part of our final spectroscopic catalogue, the full version of which is available in electronic format. Column ID is the identifier of the stars, X and Y are the positions of the stars in the VIMOS pre-images, and column r is the distance from the centre of NGC 6822 (RA = 296.24059, DEC = $-$14.80343). The other columns list the $R_0$, $J_0$, $H_0$, and $K_0$ magnitudes, the photometrically assigned spectral class \citep{sibbons+10}, the classification based on our VIMOS spectra, and in the last column the quality flag of the spectra.

\begin{table*}
\caption{Spectroscopic and photometric catalogue of stars observed in the direction towards NGC 6822. This is part of the full table, which is available in the electronic version of the paper.} \label{tab:classification}
{\scriptsize
 \begin{tabular}{ccccccccccccc}
\hline
ID & X & Y  & RA & DEC & r &R& J & H & K & Sp. Cl.& Sp. Cl.&Quality flag \\
   & [px]& [px] & [deg] & [deg] & [deg] &[mag]& [mag] & [mag] & [mag] & photometry & spectroscopy & \\
\hline
 94208& 219.732&1679.789  &  296.349976&$-$14.828187&  0.1121527&19.772& 17.877& 17.045 &16.898&   M	&      M1III	   &4\\
 94979& 330.734&1801.989  &  296.357117&$-$14.834526&  0.1206048&19.510& 17.528& 16.720 &16.515&   Mca	&      M1III	   &4\\
 96968& 608.073&1462.465  &  296.337067&$-$14.850516&  0.1073543&18.978& 16.924& 16.080 &15.693&   C	&      C5.5	   & 5\\
 97590& 696.476&1751.644  &  296.354126&$-$14.855531&  0.1249199&19.406& 17.334& 16.521 &16.218&   M	&      C5.5	   & 5\\
 99641&1001.033&2013.256  &  296.369537&$-$14.872893&  0.1464667&20.196& 17.025& 15.971 &15.317&   C	&      C6.5	   & 5\\
101911&1355.722& 158.286  &  296.260742&$-$14.892557& 0.09137722&20.645& 17.952& 17.153 &16.985&   M	&      M	   & 4\\
 77311&  48.431&1099.388  &  296.316101&$-$14.679276&  0.1453135&19.438& 17.443& 16.601 &16.179&   C	&      C6.5	   & 5\\
 79509& 475.349&2029.313  &  296.371155&$-$14.703938&  0.1641518&20.219& 18.079& 17.198 &16.991&   M	&      M1III	   &5\\
 83906&1226.993&1566.500  &  296.343781&$-$14.746560&  0.1178242&21.667& 17.437& 16.532 &16.128&   C	&      C5.5	   & 5 \\
 85530&1462.654&1902.611  &  296.363739&$-$14.759742&  0.1306686&23.344& 16.528& 15.565 &14.986&   C	&      C5.5	   & 5\\
 85885&1506.260&2089.456  &  296.374664&$-$14.762603&  0.1401522&22.981& 17.504& 16.586 &16.355&   M	&      M4III	   &5\\
 88539&1880.543&1058.628  &  296.313934&$-$14.783896& 0.07590062&22.971& 17.388& 16.359 &15.668&   C	&      C5.5	   & 4\\
 77230& 140.325&1975.243  &  296.193542&$-$14.678253&  0.1337262&20.206& 17.488& 16.498 &16.071&   C	&      C5.5	   & 5\\
 78095& 311.616&1734.144  &  296.179352&$-$14.688039&  0.1306334&19.813& 17.241& 16.350 &16.076&   M	&      C5.5	   & 5\\
 78871& 459.932&2256.131  &  296.210236&$-$14.696464&  0.1111891&21.556& 16.949& 16.337 &15.919&   M	&      dM5e	   &5\\
 79224& 532.454&1875.599  &  296.187653&$-$14.700607&  0.1156495&21.826& 17.792& 17.073 &16.777&   M	&      M6.5III     &5\\
 79511& 589.745&1521.752  &  296.166718&$-$14.703956&  0.1239035&20.233& 17.724& 16.890 &16.697&   M	&      M1III	   &4\\
 80764& 832.120&1828.010  &  296.184784&$-$14.717754&  0.1022478&20.280& 17.435& 16.459 &16.039&   C	&      C5.5	   & 5\\
 81015& 879.108&1219.159  &  296.148743&$-$14.720551&  0.1237123&20.331& 18.060& 17.254 &17.066&   M	&      M1III	   &5\\
 81290& 923.831&1041.148  &  296.138214&$-$14.723130&   0.130111&20.327& 17.227& 16.188 &15.542&   C	&      C6.5	   & 5\\
 82246&1076.064&1643.366  &  296.173828&$-$14.731837& 0.09789109&20.061& 17.335& 16.543 &16.299&   M	&      M4III	   &6\\
 83467&1267.180&2155.227  &  296.204193&$-$14.742633& 0.07085879&18.691& 15.048& 14.173 &13.874&   M	&      M6III	   &6\\
 83752&1310.298&1599.732  &  296.171265&$-$14.745203& 0.09053338&20.500& 18.020& 17.228 &16.970&   M	&      C5.5	   & 5\\
 84020&1354.477&1830.278  &  296.185242&$-$14.747572& 0.07863506&23.015& 17.622& 16.799 &16.588&   M	&      dM	   &5\\
 85975&1628.458&2046.414  &  296.197662&$-$14.763362&  0.0587216&20.414& 17.175& 16.255 &15.536&   C	&      C5.5	   & 5\\
 86798&1753.719&2238.658  &  296.208893&$-$14.769947& 0.04610623&24.466& 17.025& 16.201 &15.758&   C	&      C6.5	   & 5\\
 88152&1934.647&2076.698  &  296.199615&$-$14.780900& 0.04676038&20.389& 18.414& 17.421 &17.201&   C	&      M2III	   &5\\
 88716&2021.618&2204.011  &  296.206879&$-$14.785284& 0.03828439&19.733& 18.082& 17.209 &17.074&   M	&      C5.5	   & 5\\
 93532&  99.508&1716.032  &  296.178833&$-$14.822721& 0.06469996&20.139& 17.330& 16.475 &16.210&   M	&      SIII       & 5\\
 93858& 143.567&1738.309  &  296.180084&$-$14.825297&  0.0643363&19.755& 17.231& 16.366 &15.897&   C	&      C5.5	   & 6\\
 94446& 230.224&2278.535  &  296.212219&$-$14.830112& 0.03894694&20.872& 17.270& 16.343 &16.036&   Cca	&      C8.2	   & 5\\
 94947& 300.292&1631.935  &  296.173828&$-$14.834276& 0.07354362&20.655& 18.015& 17.151 &16.987&   Mca	&      M1III	   &5\\
 95293& 347.628&1413.000  &  296.160797&$-$14.836968& 0.08655488&20.626& 18.027& 17.370 &17.092&   Mca	&      dM3	   & 6\\
\hline
 73508\tablefootmark{1}&1469.541&1550.219  &  296.488861&$-$14.638191&  0.2982319&20.405& 17.496& 16.557 &15.974&   C     &      C6.5 v      &5\\
 68214\tablefootmark{1}& 636.571&1798.562  &  296.329102&$-$14.584000&  0.2366088& 19.231&17.611& 16.957 &16.665&   M     &      M5 III e v  &5\\
101802\tablefootmark{2}& 1328.222& 1947.256 &	296.365631& $-$14.891643&   0.1530257&17.818& 15.562& 14.7120& 14.318&	C	&    M5III	 &6\\
100545\tablefootmark{3}& 1120.211&  656.699 &	296.116089& $-$14.881053&    0.146717&20.608& 17.310& 16.2500& 15.450&	C	&    C5.5	 & 6\\
 86680\tablefootmark{3}& 1308.239& 1484.110 &	296.014832& $-$14.769041&   0.2283621&18.537& 17.050& 15.8860& 15.051&	C	&    C8.2	 & 5\\
111590\tablefootmark{3}&  852.096& 1197.049 &	295.993561& $-$14.990264&   0.3097264&19.735& 17.874& 16.7020& 15.840&	C	&    C6.5	 & 5\\
\hline
 \end{tabular}
\par}
\tablefoot{
\tablefoottext{1}{LPV defined in \citet{battinelli2011}.}
\tablefoottext{2}{IR spectrum from \citet{groenewegen2004} is available for this star.}
\tablefoottext{3}{C stars in common with the catalogue of \citet{demers2006b}.}
}
\end{table*}

\begin{figure}
\centering
\resizebox{\hsize}{!}{
\includegraphics[angle=0]{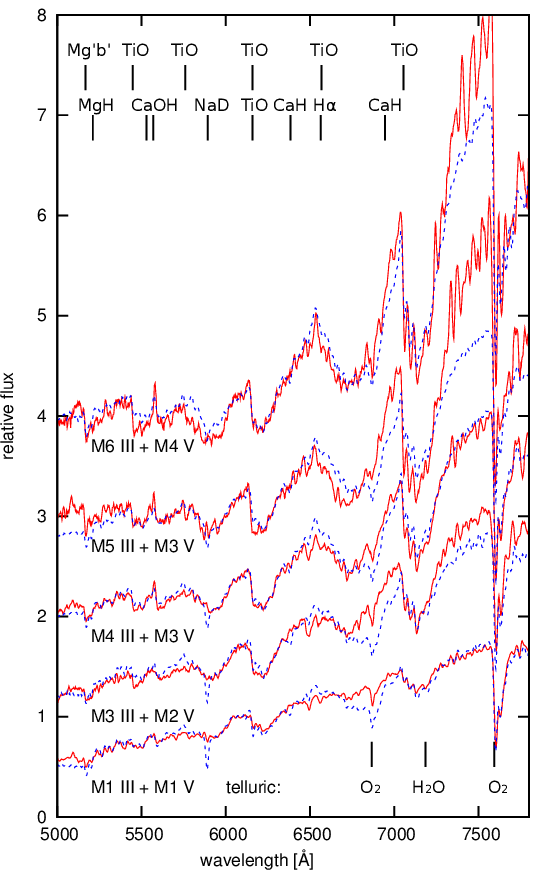}
}
\caption[]{Comparison between average dwarf and giant stars spectra with similar spectral class.
The similarities of the two types of spectra increase with increasing the spectral type. The spectra of giants are presented in red and the dwarfs with blue dotted lines. The positions of some of the most characteristic features (upper row for giants, second row for dwarfs) as well as some telluric bands (in the bottom) are shown with black lines. The spectra are shifted along the y-axis with an arbitrary constant for presentation purpose.
}
\label{fig:Dwarfs_and_Giants}
\end{figure}

\begin{figure}
\centering
\resizebox{\hsize}{!}{
\includegraphics[angle=0]{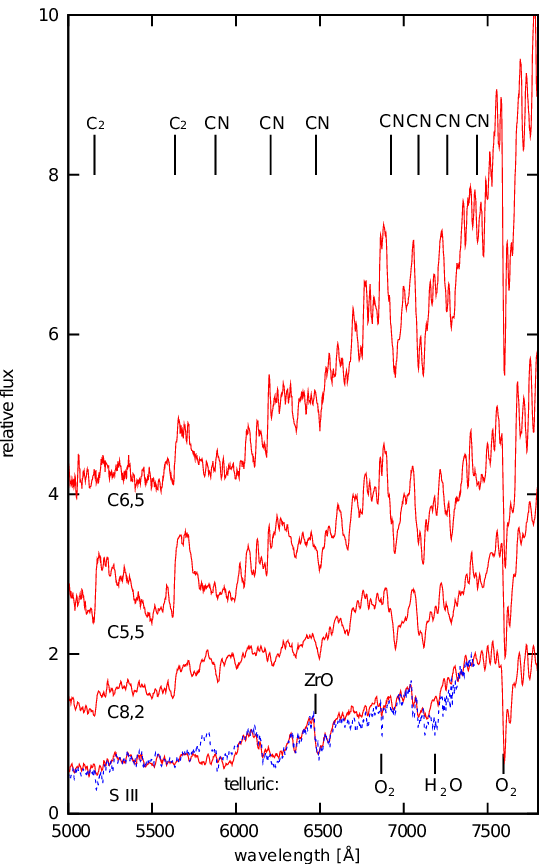}
}
\caption[]{Example spectra of individual stars of different C-types and one S-type plotted together. A spectrum of a similar star from the atlas of S stars \citep{otto+2011} is overplotted with a blue dotted line for comparison. Positions of some of the most typical features are shown with black lines: C$_2$ and CN bands for the carbon stars and ZrO for the S-type stars. The spectra are shifted along the y-axis by an arbitrary constant for presentation purpose.
}
\label{fig:CS_Stars}
\end{figure}

Different spectral subtypes are presented according to our classification in Fig. \ref{fig:Dwarfs_and_Giants}. The most similar dwarf and giant average spectra are plotted one over the other for comparison.
It is easy to follow the change of some of the spectral features characteristic of M-type stars with increasing subtype. The strong sodium doublet (blended into a single line at our resolution) is the most typical feature of luminosity class V stars. With increasing subtype, it becomes weaker and dominated by the TiO band in the same wavelength range. For the later class dwarfs, we used the CaH molecular band at $6946$\AA, which deepens and widens the TiO bands in this region, to distinguish foreground dwarfs from giants. Another typical feature of the dwarf stars is the MgH band at $5211$\AA. The very late-type M dwarfs often display H$\alpha$ in emission, which may be evidence of magnetic activity. However, H$\alpha$ was not used as a criterion for distinguishing dwarfs from giants because it may also appear in emission in long-period variable stars (usually late-type AGB stars) in certain phases owing to shock fronts \citep{lancon2000}.  
In Fig. \ref{fig:CS_Stars}, we show individual example spectra for different types of C stars, easily distinguishable by means of the many CN and C$_2$ bands and an example of a rare S-type star clearly showing ZrO bands. We also plot the spectrum of a similar galactic S star from the atlas of \citet{otto+2011} over our example for comparison. Two figures showing separately the full sample of different subtypes giant and dwarf average spectra are available in the electronic version of the paper.

In total, we determined spectral types for 511 of the 546 spectra with quality flags of 4, 5, or 6 (Table \ref{tab:statistics}). The majority turned out to be foreground dwarfs. This is in part because the colour criterion for the selection of AGB stars in NGC 6822 was on purpose quite relaxed ($(J-K)_0>0.74$~mag) to allow the selection also of early-type M stars. In addition, if there was any remaining space in the masks after the primary targets were allocated to slits, then secondary targets, most of which were foreground dwarfs, were also targeted.
Among the AGB stars belonging to NGC 6822, the majority are carbon-rich giants. Once again, this partly reflects the selection criteria, which were biased towards the AGB stars and in particular C-rich giants.

\begin{figure}
\centering
\resizebox{\hsize}{!}{
\includegraphics[angle=0]{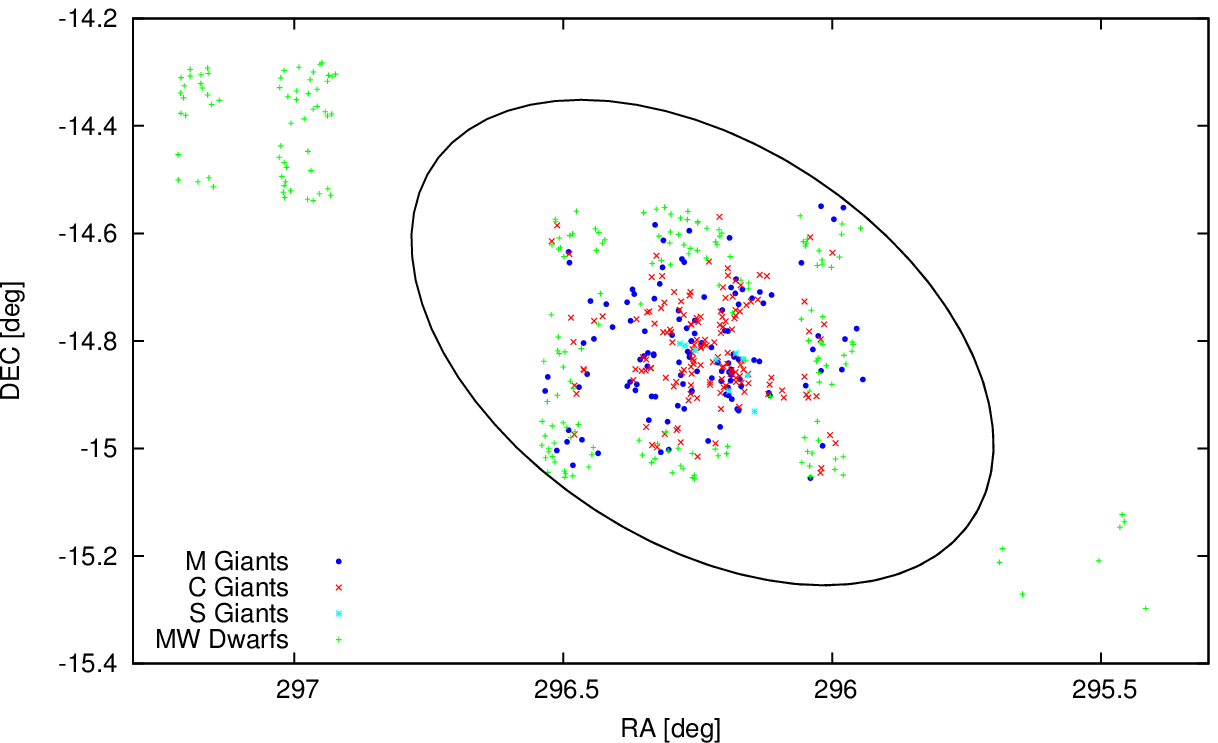}
}
\caption[]{Spatial distribution of all stars with acquired spectra of good quality. The outer $36\arcmin$ ellipse \citep{battinelli2006} is shown with solid black line. North is up and east is to the left.
}
\label{fig:Spatial_Distribution}
\end{figure}

\citet{battinelli2006} mapped the elliptical spheroid of NGC 6822 out to a semi-major axis distance of $36\arcmin$. In Fig. \ref{fig:Spatial_Distribution}, we show the spatial distribution of spectroscopically observed AGB stars in NGC 6822, as well as the Milky Way dwarfs, for which we have spectra of good quality. Essentially all giants identified by us as having either M-type or C-type AGB spectra are found within the ellipse, and the targets outside this region have been systematically found to have spectral features typical of dwarf stars.

\begin{table}
\begin{center}
\caption{Statistics of the spectral classification for the different fields.}\label{tab:statistics}
{\footnotesize
 \begin{tabular}{cccccccc}
\hline
Field & All stars & M III & C III & S III & dwarfs & Unclass. & \\
\hline
  CF  & 138 & 55 & 64 & 4 &  8 & 7 &\\
  NW1 &  47 &  7 & 10 & 0 & 28 & 2 &\\
  NE1 &  96 & 22 & 25 & 1 & 39 & 9 &\\
  SE1 & 115 & 24 & 36 & 2 & 48 & 5 &\\
  SW1 &  62 & 14 & 16 & 2 & 29 & 1 &\\
  NE2 &  70 &  0 &  0 & 0 & 62 & 8 &\\
  SW2 &  18 &  1 &  0 & 0 & 14 & 3 &\\
\hline
  All & 546 & 123 & 151 & 9 & 228 & 35 & \\
\hline
 \end{tabular}
\par}
\end{center}
\end{table}

\citet{battinelli2011} published a catalogue of 63 long-period variable stars (LPVs) in NGC 6822. Cross-correlating our spectroscopic sample with their catalogue, we found two exact matches - a C giant (ID 73508), which is an irregular variable, and an M giant semi-regular variable (ID 68214) with a $149$-day period. The latter shows H$\alpha$ line in emission. Another four variable sources were found within $10\arcsec$ of our C and S stars, but without more information about the astrometric accuracy of the catalogue, we are unable to exclude that these are spurious matches and therefore do not report them here. We also matched our catalogue with the 10 IR spectra obtained by \citet{groenewegen2009} in NGC 6822. We found two exact matches with their targeted stars, one of which they actually observed (2MASS19452775-1453299) and we confirm that it is an M-type AGB star. Finally, we matched our catalogue with the catalogue of carbon stars (142 sources) obtained by \citet{demers2006b}. We found three exact matches, which were all confirmed as C-type stars based on our spectra. These have the following IDs in \citet{demers2006b} catalogue: 1075, 1031, and 1026. Details about the stars in common with the cited catalogues are presented at the bottom of Table \ref{tab:classification} marked with different indices. All stars that were matched based on coordinates with AGB samples of other authors, were also matched in spectral classification.

We emphasize the small overlap between our spectroscopic catalogue and the cited works. This might be partially explained by the limited choice of slit positions on the VIMOS masks. We also cross-correlated our full IR catalogue with these surveys and found that it contains 33 (from 63) sources from \citet{battinelli2011}, 8 (from 10) sources from \citet{groenewegen2009}, and only 20 (from 142) sources from \citet{demers2006b}. The astrometric differences are less than $0.2\arcsec$ and the stars in common have similar $J$ and $K$ magnitudes. It is not trivial to answer where the biases in the full catalogue come from but this result might mean that our catalogue suffers from some incompleteness in either the outer regions of the galaxy, the redder part of the CMD, or both. All the stars in \citet{demers2006b} have colours $J-K > 1.5$~mag and are situated outside the central parts of NGC 6822. Other reasons could be the different criteria for selecting stellar sources or larger astrometry errors.

\section{Results and discussion}
\subsection{Colour - magnitude and colour - colour diagrams}

\begin{figure}
\centering
\resizebox{\hsize}{!}{
\includegraphics[angle=0]{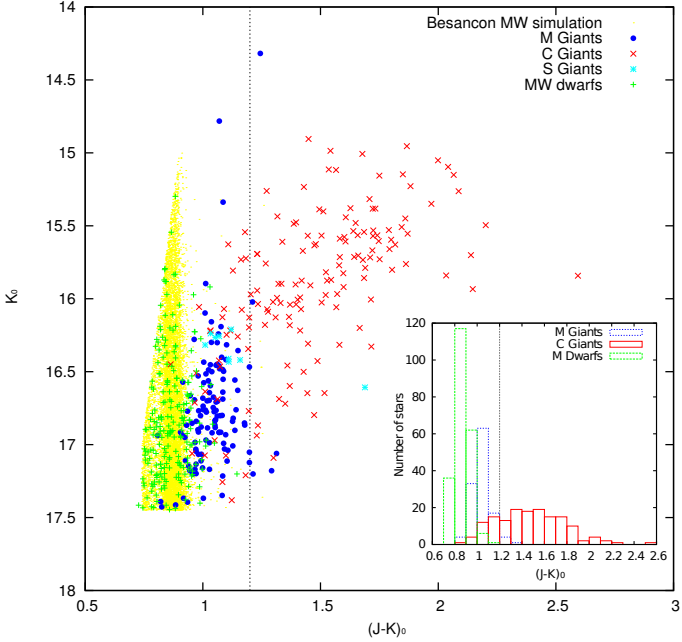}
}
\caption[]{CMD for all stars in the field of NGC 6822 with good quality spectra. The black dotted line shows the photometric selection criterion for differentiating between C- and M-type giants from \citet{sibbons+10}. The Milky Way dwarf sequence can be easily distinguished, even though it overlaps with the AGB sequence. The simulation of the foreground population obtained with the Besan\c{c}on models is indicated by yellow dots. The smaller rectangle is a histogram of this CMD.
}
\label{fig:CMD}
\end{figure}

Fig. \ref{fig:CMD} shows the $(J-K, K)$ colour-magnitude diagram (CMD) for all stars with good quality spectra. The diagram contains three distinct groups of objects - the two different types of AGB stars and the foreground population. They are depicted with different colour codes according to our spectral classification. We note that most of the M-type AGB stars have a colour index that is typically bluer than the $(J-K)_0=1.2$~mag limit \citep{sibbons+10}. The C-type stars, however, are dispersed across both sides of this limit. It is interesting that the bluer C-type stars, as well as the S-type stars are distributed preferably over the brighter end of the M star sequence. The foreground stars are distributed in a vertical sequence of $(J-K)_0\sim0.8$~mag, that is easily distinguishable but has a significant overlap with the AGB star distribution. This can be seen more easily in the histogram in Fig. \ref{fig:CMD}. The histogram presents the number of classified stars vs. the $(J-K)_0$ colour index in $0.1$~magnitude bins. Most of the foreground Milky Way dwarfs have $(J-K)_0$ colours between $0.7$ and $1.0$~mag, with maximum number of dwarfs beeing in the $0.8 - 0.9$~mag bin. Most of the M giant stars are in the $1.0 - 1.1$~mag bin and they strongly overlap with the dwarf sequence for the range $(J-K)_0 \sim 0.9 - 1.0$~mag.

We used the Besan\c{c}on model of the Milky Way \citep{robin+03} to simulate the foreground population. The model simulated about $11000$ M- and K-type main sequence stars in total, within an area of $1.14$~deg$^2$ centred on the centre of NGC 6822. The model parameters are chosen so that all simulated stars match the cuts of the full IR catalogue. The simulation agrees very well with the positions of the stars classified as dwarfs on the CMD. The simulated stars are shown with small yellow dots in Fig. \ref{fig:CMD}. The Besan\c{c}on model predicts roughly $\sim10000$ dwarfs per deg$^2$, while in the full UKIRT catalogue the density of stars is $\sim7000$ stars per deg$^2$, the large majority of which are dwarfs ($\sim6000$ per deg$^2$).

\begin{figure}
\centering
\resizebox{\hsize}{!}{
\includegraphics[angle=0]{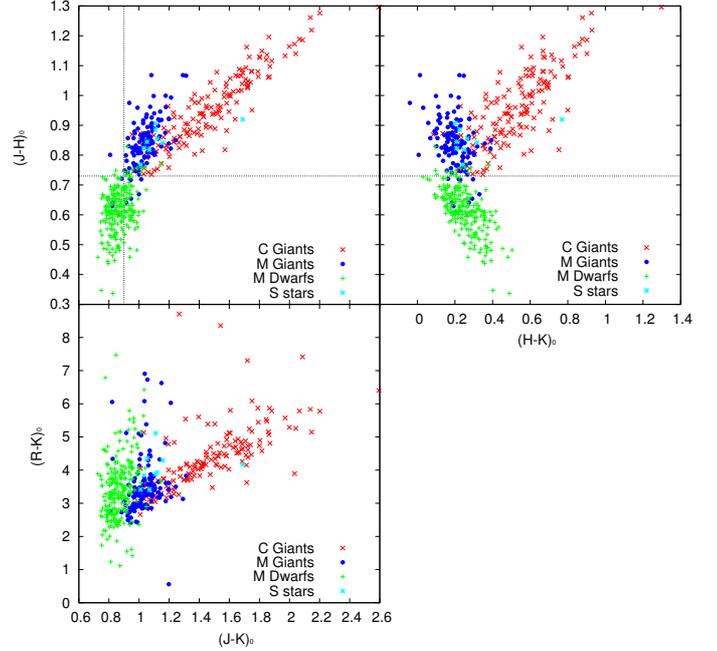}
}
\caption[]{Two-colour diagrams. Upper-left: $(J-K,~J-H)$ two-colour diagram. The two dotted lines show the AGB selection criteria: $(J-H)_0 > 0.73$~mag and $(J-K)_0 > 0.9$~mag; Upper-right: $(H-K,~J-H)$ two-colour diagram. The NGC 6822 carbon- and oxygen-rich AGB stars can be more clearly separated in this diagram; Bottom-left: $(J-K,~R-K)$ two-colour diagram.
}
\label{fig:2CD_all}
\end{figure}

Fig. \ref{fig:2CD_all} shows three two-colour diagrams. The $(H-K,~J-H)$ diagram (upper-right) is a powerful tool for separating the foreground Milky Way dwarfs and also allows a separation between the oxygen- and carbon-rich AGB stars \citep{aaronson1985, bessel1988}. In general AGB stars are expected to have $(J-H)_0>0.73$~mag and all foreground dwarf stars should be bluer than this limit \citep{gullieuszik2008, sibbons+10}. Our spectra show that this is true for all spectroscopically observed stars with few exceptions. We find that the giants bluer than the $(J-H)_0=0.73$~mag limit are mostly early M-type giants and only one is a C-type giant. There is one dwarf star, which lies significantly redward of the colour limit at $(J-H)_0 = 0.82$~mag. There is also a possibility that, owing to the high density of stars, few slits captured the light from neighbouring dwarf stars that were not the main target. We found this to be the case for one photometrically classified C-type star, which was targeted in two masks. In the spectrum from the first mask, we can clearly see a carbon-rich star and in the spectrum from the second mask, a foreground K dwarf. We excluded the spectra of four dwarf stars, which had been photometrically classified as carbon rich-stars, because of this effect. 
In the $(J-K,~J-H)$ two-colour diagram (upper-left), the C- and M-type giants are more intermixed but it shows our selection criteria for AGB stars. The horizontal line at $(J-H)_0=0.73$~mag separates the foreground from NGC 6822 stars, and the vertical line at $(J-K)_0=0.9$~mag indicates the blue limit for the AGB stars.

We also present the $(J-K,~R-K)$ two-colour diagram (bottom-left) in Fig. \ref{fig:2CD_all}. It shows a clear separation between the different types of stars, which is however not as clean as for the IR photometry. This is mostly due to the larger R-band photometry errors and the probable variability of the targeted AGB stars (for which IR and R-band photometry were taken in different epochs), which would smear out their locations in the two-colour diagram. \citet{otto+2011} indicate that S-type stars are more clearly separated in this visible-IR two-colour diagram, which is also the case here. The colours of the Galactic S stars from \citet{otto+2011} are similar to the colours of S stars in NGC 6822.


\subsection{Photometric criteria for C vs. M giants selection based on the spectroscopic sample}

On the basis of our spectroscopic sample, we derive new photometric selection criteria for distinguishing between different types of AGB stars in NGC 6822 and foreground population. As already mentioned, the photometry is based on the near-IR catalogue of \citet{sibbons+10}. The selected photometric sample includes over $21000$ stars found within a radius of $1$~degree from the centre of NGC 6822, that are brighter than $K_0 = 17.45$~mag in accordance with the value for the TRGB of NGC 6822 from \citet{sibbons+10}.

We find that the easiest way to differentiate between the AGB population of NGC 6822 and the foreground dwarfs is to use these two-colour diagrams. We tested several $(J-H)_0$ cuts, and our results are summarized in Table \ref{tab:AGB}. This table indicates the percentage of spectroscopically confirmed AGB stars that would be identified as AGB stars using a certain cut in $(J-H)_0$. We note that our spectroscopic sample contains almost no stars with the colours $(J-H)_0 > 0.73$~mag and $(J-K)_0 < 0.9$~mag (Fig. \ref{fig:2CD_all} upper-left diagram). These represent about $3\%$ of the whole photometric sample, and because we are unable to determine their type, we exclude them. In this work, we use $(J-H)_0 > 0.73$~mag and $(J-K)_0>0.9$~mag to select AGB stars and $(J-H)_0<0.73$~mag to select foreground stars. After applying these cuts, we found that our photometric sample includes about $2800$ AGB candidates.

\begin{table}
\begin{center}
\caption{AGB stars selection from the photometric catalogue.}\label{tab:AGB}
{\footnotesize
 \begin{tabular}{cccccc}
\hline
Criterion &  M III & C III & S III & dwarfs   &\\
\hline
$(J-H)_0>0.72$ & $93\%$ & $99\%$ & $100\%$ & $9\%$ &\\
$(J-H)_0>0.73$ & $93\%$ & $99\%$ & $100\%$ & $6\%$ &\\
$(J-H)_0>0.74$ & $92\%$ & $98\%$ & $100\%$ & $5\%$ &\\
$(J-H)_0>0.75$ & $88\%$ & $96\%$ & $100\%$ & $3\%$ &\\
\hline
$(J-H)_0>0.73$~and & $92\%$ & $99\%$ & $100\%$ & $5\%$ &\\
$(J-K)_0>0.90$&&&&&\\
\hline
\end{tabular}
\par}
\end{center}
\end{table}

\begin{figure}
\centering
\resizebox{\hsize}{!}{
\includegraphics[angle=0]{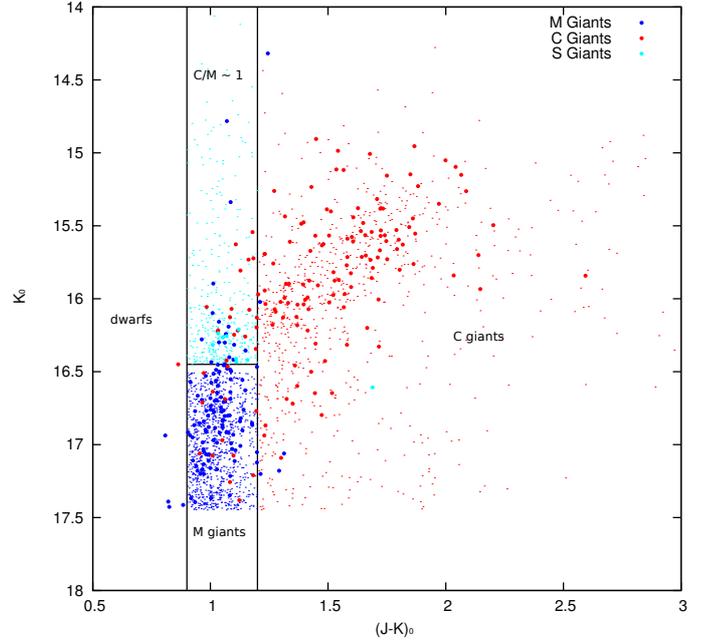}
}
\caption[]{First C- and M-type set of selection criteria. Most of the S-type stars and an equal number of C- and M-type stars are found in the region $K_0<16.45$~mag, $(J-K)_0<1.2$~mag. Excluding this region, our selection is based on a colour cut at $(J-K)_0=1.2$~mag. The photometric sample is presented with dots and the spectroscopically confirmed stars with larger points.
}
\label{fig:crit_I}
\end{figure}

We propose two different approaches for distinguishing between carbon- and oxygen-rich AGB stars from the $(J-K,~K)$ CMD in order to estimate the C/M ratio. As we mentioned before, almost all S-type stars (8 of 9) and most of the C-type stars with colours $(J-K)_0<1.2$~mag are brighter than $K_0 = 16.45$~mag. We can also see that there are roughly equal numbers of M- and C-type stars in the region $(J-K)_0<1.2$~mag and $K_0<16.45$~mag according to our spectroscopic sample: 17 M-type and 18 C-type stars. Our first set of selection criteria (Fig. \ref{fig:crit_I}) is based on this result. We exclude all objects for which $(J-K)_0<1.2$~mag and $K_0<16.45$~mag. This permits us to remove the S-type stars as well as the region in the CMD where the C- and M-type stars strongly overlap but have a number ratio close to $1$. We then simply divide the CMD into two parts: stars bluer than $(J-K)_0=1.2$~mag, which we call M giants, and redder than $(J-K)_0=1.2$, which are C giants. If we adopt a distance modulus of $23.35$~mag (weighted average obtained from Tables 4 and 5 in \citet{gorski2011}), the absolute K magnitude of this limit will be $M_K = -6.90$~mag.

\begin{figure}
\centering
\resizebox{\hsize}{!}{
\includegraphics[angle=0]{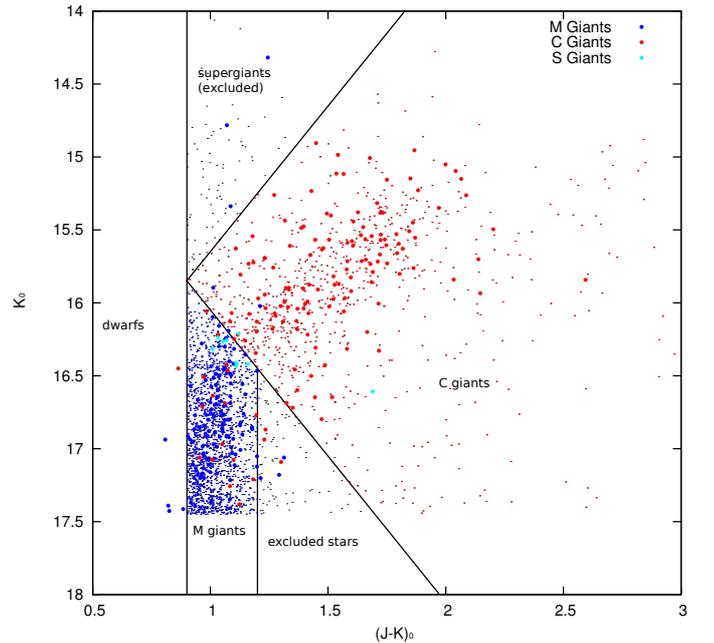}
}
\caption[]{Second C- and M-type set of selection criteria. The two excluded regions contain only $6\%$ of the AGB candidates. The photometric sample is presented with dots and the spectroscopically confirmed stars with larger points.
}
\label{fig:crit_II}
\end{figure}

The second approach is also based on cuts in the $(J-K,~K)$ CMD. We assume that all objects with $(J-K)_0<1.2$~mag and $K_0>2\times(J-K)_0+14.05$ are M-type giants and objects with $K_0<2\times(J-K)_0+14.05$ and $K_0>-2\times(J-K)_0+17.65$ are C-type giants (Fig. \ref{fig:crit_II}). 
This method provides the clearest possible separation between carbon- and oxygen-rich AGB stars but does not account for S-type stars. The excluded regions contain only $6\%$ of the AGB candidates and the upper region might consist mostly of supergiants and foreground stars. In the fainter excluded region, we have three spectroscopically confirmed M giants and three C giants. Although we recognize the effect of small number statistics, there appears to be a large overlap between the C- and M-type stars in this region of the CMD. Adopting a distance modulus of $23.35$~mag, and correcting for it, the selection criteria are all objects with $(J-K)_0<1.2$~mag and $M_K>2\times(J-K)_0-9.30$ are M-type giants and objects with $M_K<2\times(J-K)_0-9.30$ and $M_K>-2\times(J-K)_0-5.70$ are C-type giants.

Tables \ref{tab:CMAGB} and \ref{tab:cont} present the statistics of our selection criteria. In particular, Table \ref{tab:CMAGB} illustrates the effectiveness of both selection criteria. It lists the percentage of all spectroscopically classified stars for each spectroscopic class (e.g. M III, C III, S III) that coincides with the various photometric selection boxes (M-type selection, C-type selection, excluded). Table \ref{tab:cont} instead shows the expected contamination of the M-type and C-type photometric selection boxes. This is estimated by assuming that the total number of spectroscopically confirmed stars in a given selection box (e.g. M-type selection or C-type selection box) is $100\%$ and then computing the percentage of M-type, C-type, S-type, or dwarfs in that sample. We refer to this table when we discuss the overall C/M ratio in Sect. 4.3.
The two selection criteria are good for the purpose of deriving the C/M ratio because some regions of the CMD, with reciprocal contamination, are excluded. They cannot be used to obtain the absolute number of C- and M-type stars in the galaxy.

\begin{table}
\begin{center}
\caption{Results of the C- and M-type star selection criteria.}\label{tab:CMAGB}
{\footnotesize
\begin{tabular}{cccc}
\hline
I set criteria: & M-type sel.          & C-type sel.        &  Excluded \\
\hline
Sp. M-type      & $74\%+(14\%)$   & $4\%$          &  $(14\%)$   \\
Sp. C-type      & $9\%$           & $79\%+(12\%)$  &  $(12\%)$   \\
Sp. S-type      & $0\%$           & $11\%$         &  $89\%$     \\
\hline
II set criteria:&&&\\
\hline
Sp. M-type      & $84\%$          & $2\%$          &  $6\%$      \\
Sp. C-type      & $11\%$          & $86\%$         &  $2\%$      \\
Sp. S-type      & $78\%$          & $22\%$         &  $0\%$      \\
\hline
\end{tabular}
\par}
\end{center}
\end{table}

\begin{table}
\begin{center}
\caption{Expected contamination by different types of stars in the selection boxes.}\label{tab:cont}
{\footnotesize
 \begin{tabular}{ccc}
\hline
I set criteria: & M-type selection & C-type selection \\
\hline
M-type cont.     & $79\%$          & $4\%$         \\
C-type cont.     & $11\%$          & $95\%$        \\
S-type cont.     & $0\%$           & $1\%$         \\
dwarfs cont.     & $10\%$          & $0\%$         \\
\hline
II set criteria:&&\\
\hline
M-type cont.     & $75\%$          & $2\%$          \\
C-type cont.     & $12\%$          & $97\%$         \\
S-type cont.     & $5\%$          & $1\%$         \\
dwarfs cont.     & $8\%$          & $0\%$         \\
\hline
\end{tabular}
\par}
\end{center}
\end{table}

\subsection{C/M ratio and metallicity distribution}

\begin{figure*}
\centering
\resizebox{\hsize}{!}{
\includegraphics[angle=0]{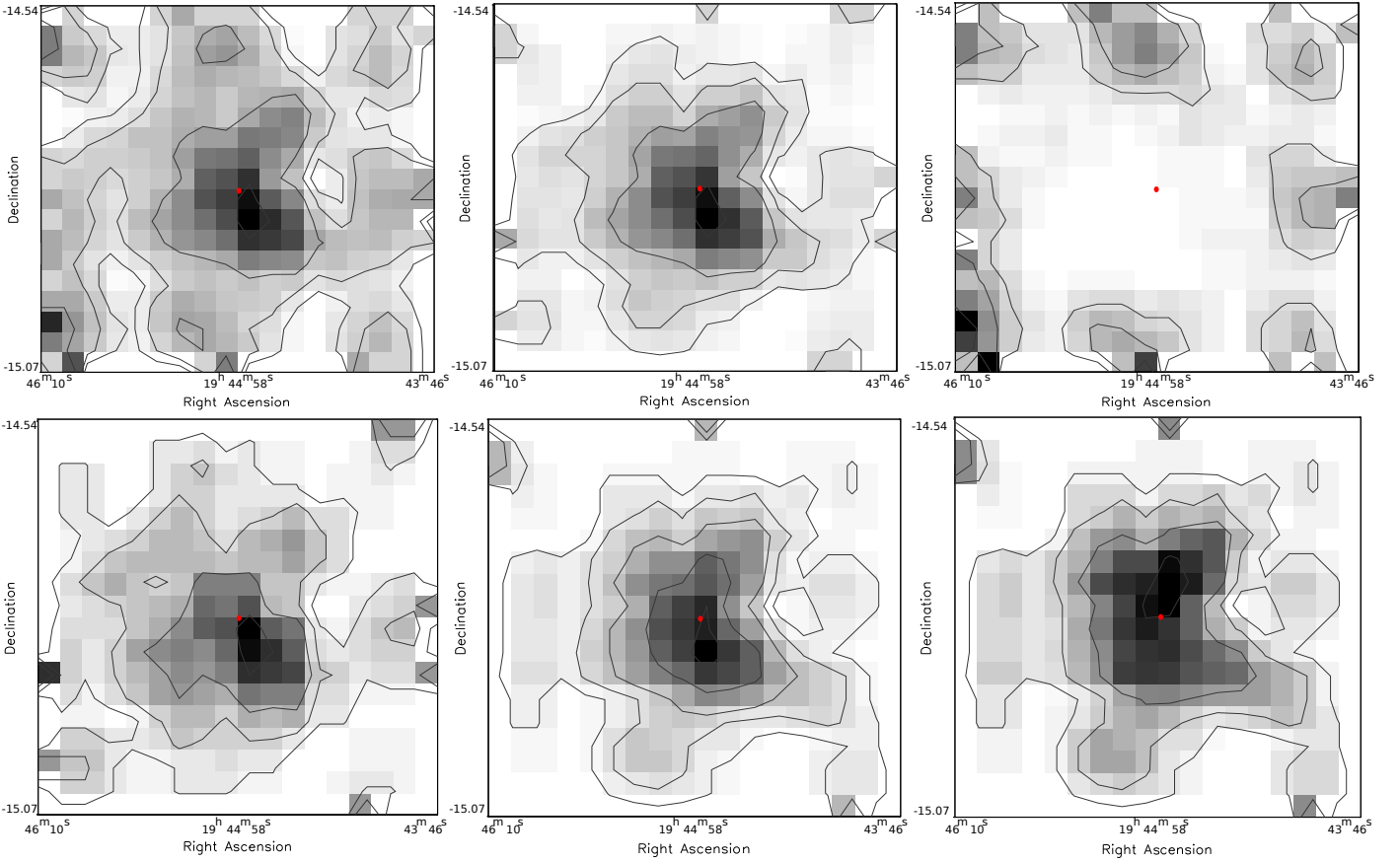}
}
\caption[]{Logarithmic and smoothed density distributions of all stars with good quality spectra. The maps are in $18\times 18$ bins $2.0\arcmin\times 1.8\arcmin$ each as follows; top: all stars, all AGB stars, and all dwarf stars (contours are for $0.5, 1, 2, 5$ density levels); bottom: M-type AGB stars, C-type AGB stars (contours are for $0.2,0.5,1,2,3$ density levels), and C/M ratio map (contours are for ratios $0.2,0.5,1,2$). Darker regions correspond to higher density. The centre of the galaxy is indicated with a small red dot. North is up and east is to the left.
}
\label{fig:Maps}
\end{figure*}

\begin{figure*}
\centering
\resizebox{\hsize}{!}{
\includegraphics[angle=0]{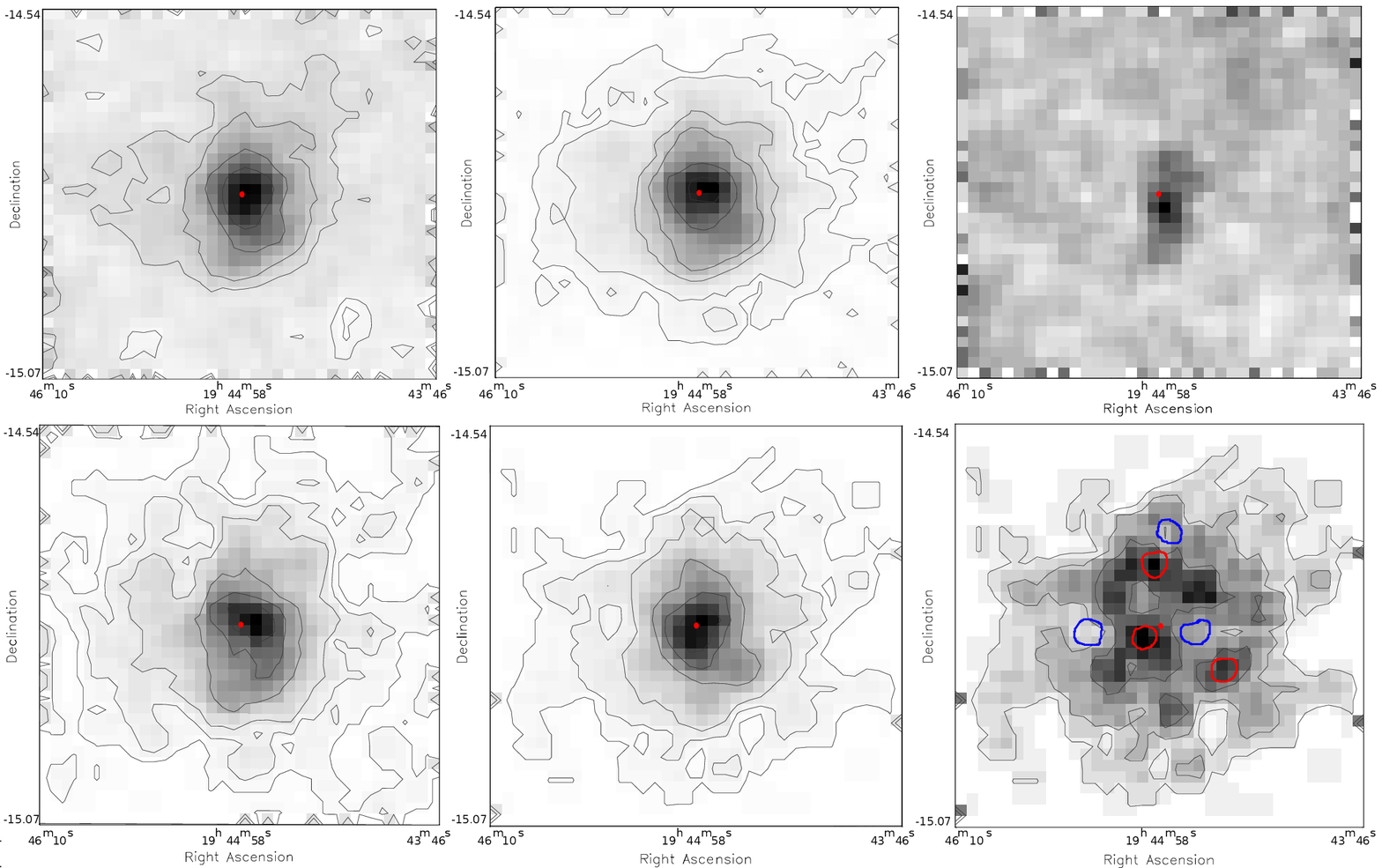}
}
\caption[]{Logarithmic and smoothed density distributions of the stars from the near-IR catalogue. The maps are in $36\times 36$ bins $1.0\arcmin\times 0.9\arcmin$ each as follows; top: all stars, all AGB stars, and all dwarf stars (contours are for $0.5, 1, 3, 5, 10, 15$ density levels); bottom: M-type AGB stars, C-type AGB stars (contours are for $0.2,0.5,1,3,5,7$ density levels), and C/M ratio map (contours are for ratios $0.2,0.5,1$). Darker regions correspond to higher density. The centre of the galaxy is indicated with a small red dot. The blue circles indicate low and the red circles high C/M regions that are further discussed in Sects. 4.4 and Fig. \ref{fig:isochrones}. North is up and east is to the left.
}
\label{fig:Maps_phot}
\end{figure*}

Fig. \ref{fig:Maps} shows the number density of sources with good quality spectra. We counted the stars of a given spectral type in $18\times 18$ bins, where a single bin corresponds to $2.0\arcmin\times 1.8\arcmin$ rectangle ($36\arcmin\times20\arcmin$ field). The source density in each map is smoothed with a boxcar function of $width = 2$ prior to the construction of the greyscale images, where higher concentrations of sources are indicated with darker regions. From left to right, we show all stars, all AGB stars, and the foreground stars in the first row, and M-type AGB stars, C-type AGB stars, and the C/M ratio map in the second row. The size of the maps corresponds to the observed field within the outer ellipse of NGC 6822 according to Fig. \ref{fig:Spatial_Distribution} and is based on $271$ AGB stars in total: $121$ M- and $150$ C-type giants. The number of individual stars is smaller than the number of obtained spectra because few stars were targeted in two of our masks. We can see from Fig. \ref{fig:Maps} that the AGB population is concentrated at the centre of the galaxy. However the large C/M ratio in the central parts (which varies between $1$ and $2$) is due to underestimating the number of M stars with respect to the C stars and it is possible that the small C/M ratio in the outer parts is due to small number statistics. These maps reflect the biases in the spectroscopic sample (C stars were targeted preferentially) and cannot be used to derive the metallicity distribution within the galaxy. 

To investigate the C/M ratio and metallicity distribution within NGC 6822, we use the complete near-IR photometric catalogue. Here we apply the second set of photometric selection criteria (discussed in the previous section) to distinguish between carbon- and oxygen-rich AGB stars. We obtain very similar results when the first set of criteria is used. The results are presented in Fig. \ref{fig:Maps_phot}, which is almost identical to Fig. \ref{fig:Maps} but the stars are binned in $36\times 36$ bins, of $1\arcmin\times 0.9\arcmin$ each. The map is based on $1753$ AGB stars in total, $970$ and $783$ photometrically classified M- and C-type stars, respectively. We would expect the foreground density map to be flat but we can see some overdensity of stars at the centre coinciding with the galaxy. This means that we underestimate the number of AGB stars, which are most probably M-type giants. This overdensity is quite small, only $1.3~\sigma$ over the foreground fluctuations and has no significant effect on our C/M ratio estimate.

The overall C/M ratio is $0.8$ according to the number of stars in the selection boxes. However, if we take into account the expected contaminations in the selection boxes from Table \ref{tab:cont}, i.e. the number of M giants to be $75\%$ of the number of stars in the M-type selection box plus $2\%$ of the number of stars in the C-type selection box plus $4\%$ from the dwarfs selection and the number of C giants to be $97\%$ of the C-type selection plus $12\%$ of the M-type selection, we calculate a C/M ratio $\sim1.05$. This value is in excellent agreement with C/M ratio of $1.0$ obtained by \citet{letarte2002}.

We see a similar trend in the C/M ratio distribution across the galaxy as described by \citet{cioni+habing05}. The galaxy centre has a relatively small C/M ratio that increases when moving outwards. The maximum of the C/M ratio follows a broken ellipse around the galactic centre and then starts to decrease with increasing radii. This decrease in the C/M ratio towards the outer regions may be real or due to small number of statistics. A bar-like structure traced by younger stars in \citet{gouliermis2010} cannot be distinguished in the C/M ratio distribution. The area investigated by \citet{karampelas2009} corresponds to the field of Fig. \ref{fig:Maps_phot}. Beyond that, the density of stars is too low and it is difficult to estimate the C/M ratio. We discuss the stellar density in the outermost regions of our photometric catalogue in Sect. 4.5.

The variation in the C/M ratio across the face of the galaxy can be explained by a variation in the metallicity. A higher C/M ratio relates to a lower metallicity. This relation is well-studied in previous works of \citet{cioni+habing03,cioni+habing05} for the SMC, LMC, M 33, and this galaxy, and calibrations between C/M and [Fe/H] were presented in \citet{groenewegen2004}, \citet{battinelli2005}, and \citet{cioni2009b}. The physical reasons for this correlation are explained by \citet{scalo1981} and \citet{iben1983}: (i) O-rich AGB stars of lower metallicity turn more easily into C-rich stars; (ii) evolutionary tracks for lower metallicities correspond to higher temperatures; (iii) in very low metallicity environments, post-horizontal branch stars may fail to become AGB stars.

In our study of NGC 6822, the C/M ratio varies between $0.2$ and $1.8$ according to Fig. \ref{fig:Maps_phot} and this corresponds to a metallicity variation of $\Delta\mathrm{[Fe/H]}\sim0.4$~dex (between $-0.9$~dex and $-1.3$~dex) with an average $\mathrm{[Fe/H]}\sim-1.2$~dex using the calibration proposed by \citet{groenewegen2004}

\begin{equation}
\mathrm{[Fe/H]} = -0.42 \times \log \mathrm{(C/M)} - 1.23. 
\end{equation}
Another calibration proposed by \citet{battinelli2005} is
\begin{equation}
\mathrm{[Fe/H]} = -0.59 \times \log \mathrm{(C/M)} - 1.32.
\end{equation}
Using this we obtain an average value of $\mathrm{[Fe/H]}\sim-1.3$~dex with a spread of between $-0.9$~dex and $-1.5$~dex.
A revision of Eq. 4 is presented in \citet{cioni2009b}, who propose the alternative relation
\begin{equation}
\mathrm{[Fe/H]} = -0.47 \times \log \mathrm{(C/M)} - 1.39.
\end{equation}
For NGC 6822, we get an average value of the [Fe/H] index $\sim -1.3$~dex with a spread of between $-1.1$~dex and $-1.5$~dex.
If we adopt C/M ratio $\sim 1.05$, the obtained metallicity values are slightly lower, but still very similar to the [Fe/H] values discussed here.
The metallicity estimates are in good agreement with the results of \citet{tolstoy2001} and \citet{davidge2003}. The former study found an average [Fe/H] $=-1.0\pm0.5$~dex with a spread of between $-0.5$ and $-2.0$~dex based on Ca II spectroscopic measurements of 23 RGB stars, while the latter derived [Fe/H] $=-1.0\pm0.3$~dex from the slope of the RGB.
A measurement of the Ca II triplet absorption lines in a statistically significant number of AGB stars in NGC 6822 is the subject of a subsequent paper by our team to provide an independent estimate of the metallicity index.



\subsection{C/M ratio and possible age variations}

\begin{figure}
\centering
\resizebox{\hsize}{!}{
\includegraphics[angle=0]{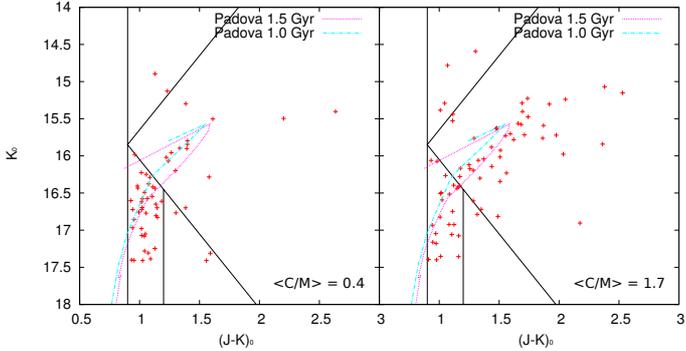}
}
\caption[]{Stars from three regions of high and low C/M ratios are plotted and compared with isochrones for [Fe/H] $= -1.0$~dex and ages $1.0$ and $1.5$~Gyr. The selected regions are indicated in Fig. \ref{fig:Maps_phot}. It seems that larger ratios refer to slightly older ages.
}
\label{fig:isochrones}
\end{figure}

\citet{feast+2010} suggested that radial trends of the C/M ratio across the galaxy might also be explained by age rather than metallicity variations.
We test this possibility by fitting isochrones from the Padova library \citep{marigo+08,girardi+10}. The comparison with isochrones shows that the AGB stars are older than $0.8$ Gyr but an upper age limit is difficult to establish because the isochrones become uncertain for ages $>2.0$~Gyr in this region of the CMD. We compare different regions of high and low C/M ratios in the galaxy with isochrones for ages $1.0$ and $1.5$~Gyr and metallicities between $-0.7$ and $-1.3$~dex. We see that regions with larger ratios have slightly tighter fits by isochrones of age $1.5$~Gyr, thus are a bit older than the regions with relatively smaller ratios, which are more consistent with isochrones of age $1.0$~Gyr (Fig. \ref{fig:isochrones}). The selected regions are marked on the C/M ratio map in Fig. \ref{fig:Maps_phot}: blue for smaller and red for larger ratios. We note, however, that there are a number of uncertainties such as whether there is a significant intrinsic reddening, and any uncertainty in both the distance modulus and the theoretical models themselves, that could bias the conclusions.

In accordance with this line of thought, the slight shift in the peak of the M star distribution with respect to the C stars is unsurprising and probably means that the photometric selection criteria select M stars that are younger than the bulk of C stars. While the densest concentration of C stars is located at the centre of NGC 6822, the M star distribution peaks about $1\arcmin$ west of it, centred on a bright UV-emission region of recent star formation detected by GALEX (region 26, defined in \citet{efremova+2011}).

\subsection{Structure of NGC 6822}

\begin{figure}
\centering
\resizebox{\hsize}{!}{
\includegraphics[angle=0]{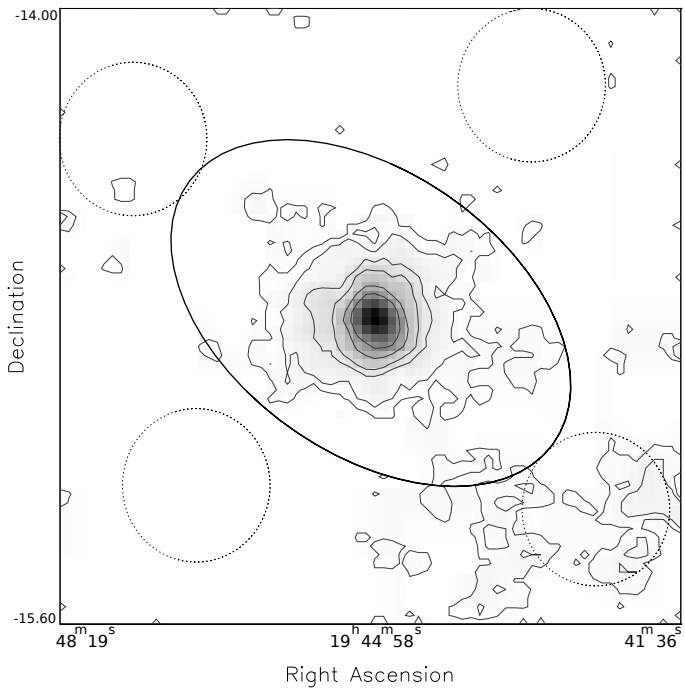}
}
\caption[]{Smoothed and logarithmic density map of all AGB candidates within the entire field of the IR catalogue. We see a rather significant excess of stars south-west of the centre of the galaxy. The map consists of $72\times72$~bins, $1.4\times1.3$~arcmin each. Contours are for 0.5, 1, 3, 5, 10, 15 density levels. The \citet{battinelli2005} ellipse and the four discussed outer fields are sketched along. North is up and east is to the left.
}
\label{fig:AGB_excess}
\end{figure}

\begin{table*}
\begin{center}
\caption{Stellar density in four outer fields around NGC 6822, considering only the AGB population ($K<17.45$~mag).}\label{tab:outside}
{\footnotesize
 \begin{tabular}{cccccccccccc}
\hline
Field & RA & Dec & Radius & M III & C III & Dwarfs &Fraction of AGB& Total st. density    & $E(B-V)$ &\\
      & [deg]& [deg] & [deg]    & count & count & count  &AGB/all& [stars p. sq. arcmin]  & [mag]      &\\
\hline
NW   & 295.794 & $-$14.198 & 0.2 & 25 (48) & 0 & 892 &0.03& 2.028 & 0.172 &\\
SE   & 296.702 & $-$15.240 & 0.2 & 13 (13)  & 3 & 848 &0.02& 1.911 & 0.226 &\\
NE   & 296.873 & $-$14.338 & 0.2 & 31 (43) & 2 & 746 &0.04& 1.723 & 0.187 &\\
SW   & 295.620 & $-$15.302 & 0.2 & 130 (130) & 11& 864 &0.14& 2.223 & 0.241 &\\
\hline
\end{tabular}
\par}
\end{center}
\end{table*}

The near-IR catalogue covers a region of roughly one degree from the centre of NGC 6822. This allows us to investigate the stellar density of probable galaxy members to this extent. Using the second set of selection criteria, we found that $93\%$ of the photometrically classified C giants and only $64\%$ of the M giants are within the outer ellipse of \citet{battinelli2005}. We studied the stellar density and the interstellar extinction (using the \citet{schlegel+98} extinction maps) towards four fields (indicated in Fig. \ref{fig:AGB_excess} and referred to as NW, SW, NE and SE in Table \ref{tab:outside}) outside the Battinelli ellipse. We note that a larger number of possible galaxy members extends to the south-west at least one~degree from the NGC 6822 centre. The overall stellar density in the south-west field is indeed a little higher than the average foreground density, and the extinction is the same as towards the centre of the galaxy, $E(B-V) = 0.24$~mag.

We see that there is some variation in the extinction between these four fields, which is lower in the north-west and north-east fields, but that this difference corresponds to a $\Delta E(J-H) = 0.02$~mag correction to the $J-H$ colour index, which is the principal means of AGB star selection. Hence, we can assume that the IR extinction in the field of the entire near-IR catalogue is rather constant. In the M III column of Table \ref{tab:outside}, we list the number of M-type AGB candidates per field assuming a constant $E(B-V) = 0.24$~mag in all fields and in brackets we list the number of M-type AGB candidates that we would have if we adopt the appropriate extinction value from the last column of Table \ref{tab:outside}. The extinction correction does not lead to any new C-type AGB candidates in the sample but we note that a small correction to the $J-H$ limit could have a significant impact on the number of M-type AGB stars and foreground stars. About $10\%$ of the photometrically recognized galaxy members may actually be foreground dwarfs according to our selection criteria. A logarithmic and smoothed density map of all AGB candidates within the field of the near-IR catalogue is presented in Fig. \ref{fig:AGB_excess}. The isodensity contours that we obtain are very similar to the map presented in \citet[see their Fig. 9]{letarte2002}.

We can also consider the full photometric catalogue, which, although not complete, extends to a faintness level of $K_0=19$~mag. We adopt $(J-H)_0>0.73$~mag to select RGB and AGB candidates (a $J-K$ limit is not adopted because of the expected slope in the RGB) and investigate the stellar densities within the same four outer fields.
We again see higher density of possible members of NGC 6822 towards the south-west, although less pronounced because of the unavoidable higher contamination of foreground stars in the selected sample.

Hence, this excess of stars seems to be real but a spectroscopic confirmation of their membership to the galaxy is required.


\section{Conclusion}

We have presented spectroscopic observations of about 800 stars in the field of the NGC 6822 dwarf galaxy, 511 of which are of good enough quality to permit a reliable spectral classification. The observed stars have been classified as M, C, and S spectral types or foreground dwarfs according to their typical spectral features. We have presented the largest spectroscopic catalogue to date of carbon and oxygen rich AGB stars in NGC 6822. Their distribution in colour-magnitude and colour-colour diagrams was discussed, and we proposed and quantified new photometric selection criteria between the different types of stars. Foreground stars tend to have colours $(J-H)_0<0.73$~mag and follow a vertical sequence with a peak at $(J-K)_0\sim0.8$~mag on the $(J-K)_0$ vs. $K_0$ CMD. This was also confirmed by the Besan\c{c}on Milky Way model \citep{robin+03}. We found, however, that this model overpredicts the stellar density in the direction of NGC 6822. We also conclude that a small variation in the dwarfs vs. AGB selection criteria may have a significant impact on the number of AGB stars with respect to the foreground.

Our selection of C- and M-types giants was based on selection boxes in the $(J-K)_0$ vs. $K_0$ CMD. These criteria were applied to the near-IR photometric catalogue of \citet{sibbons+10} and the surface distribution of the C/M ratio was discussed. We used the C/M ratio as a metallicity indicator and found that the galaxy has an average metallicity index $\mathrm{[Fe/H]}\sim-1.2\div-1.3$~dex with a spread of $\Delta\mathrm{[Fe/H]}\sim0.4\div0.6$~dex according to the different C/M vs. metallicity calibrations. Regions of larger C/M ratio (lower metallicity) are preferably distributed across a broken ellipse around the centre, which itself has a smaller ratio (higher metallicity). We also discussed whether the trends in the C/M ratio are driven by age rather than metallicity variations. A comparison with isochrones suggests that regions of higher C/M ratio are slightly older than the regions of lower C/M ratio.

The wide area of the IR catalogue allows us to investigate the field of NGC 6822 out to one~degree from its centre. We detected a significant overdensity of possible AGB stars south-west of the galaxy centre, but confirmation of their membership would require spectroscopic data.


\begin{acknowledgements}
The spectroscopic observations were acquired together with L. Sibbons, whose assistance is gratefully acknowledged. We are indebted to Carlo Izzo for his dedicated support and development of the VIMOS pipeline, including comprehensive documentation and helpful suggestions on data reduction. We thank Antoniya Valcheva for useful comments and notes on the paper. We gratefully acknowledge funding for this project by the ESO DGDF grant. The project was partially funded by the Bulgarian NSF (contract No DDVU02/40/2010). This research has made use of the NASA/IPAC Extragalactic Database (NED) which is operated by the Jet Propulsion Laboratory, California Institute of Technology, under contract with the National Aeronautics and Space Administration. We thank the anonymous referee and the editor Ralf Napiwotzki, whose comments helped to improve the paper. 
\end{acknowledgements}

\bibliographystyle{aa}

\bibliography{mybiblio_v2}

\Online

\begin{appendix}

\section{Full sample of M-type average spectra for giant and dwarf stars}

\begin{figure}
\centering
\resizebox{\hsize}{!}{
\includegraphics[angle=0]{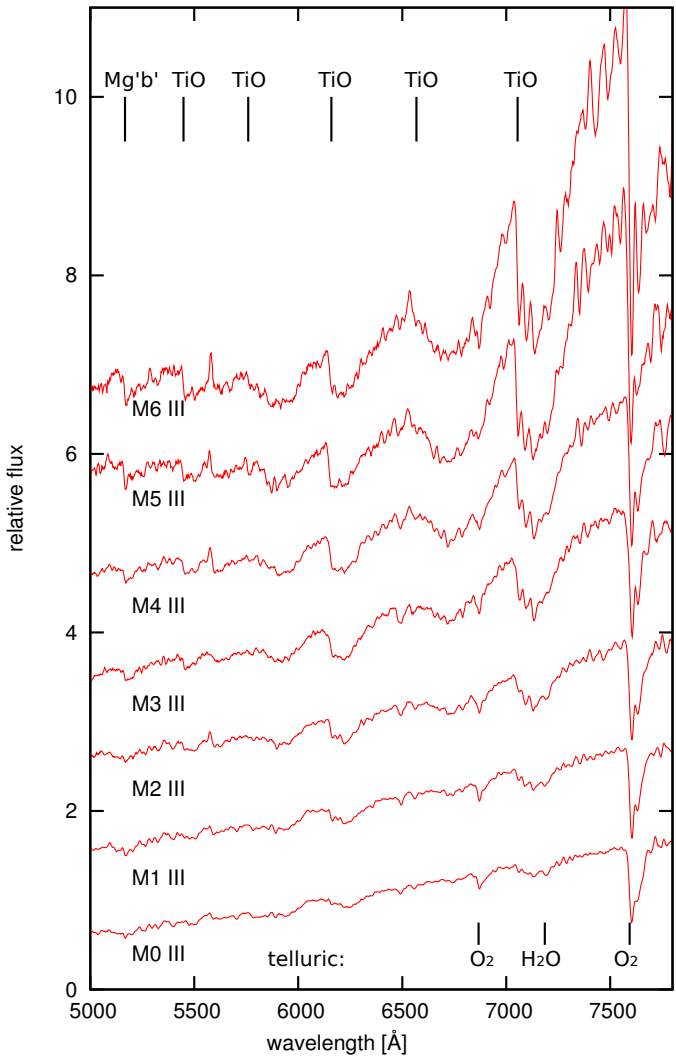}
}
\caption[]{Full sample of average spectra for different spectral types giants according to our classification. The blue edge positions of some of the strongest TiO bands are shown with black lines as well as some telluric bands. The spectra are shifted along the y-axis with an arbitrary constant for presentation purposes.
}
\label{fig:MIII_Averages}
\end{figure}

\begin{figure}
\centering
\resizebox{\hsize}{!}{
\includegraphics[angle=0]{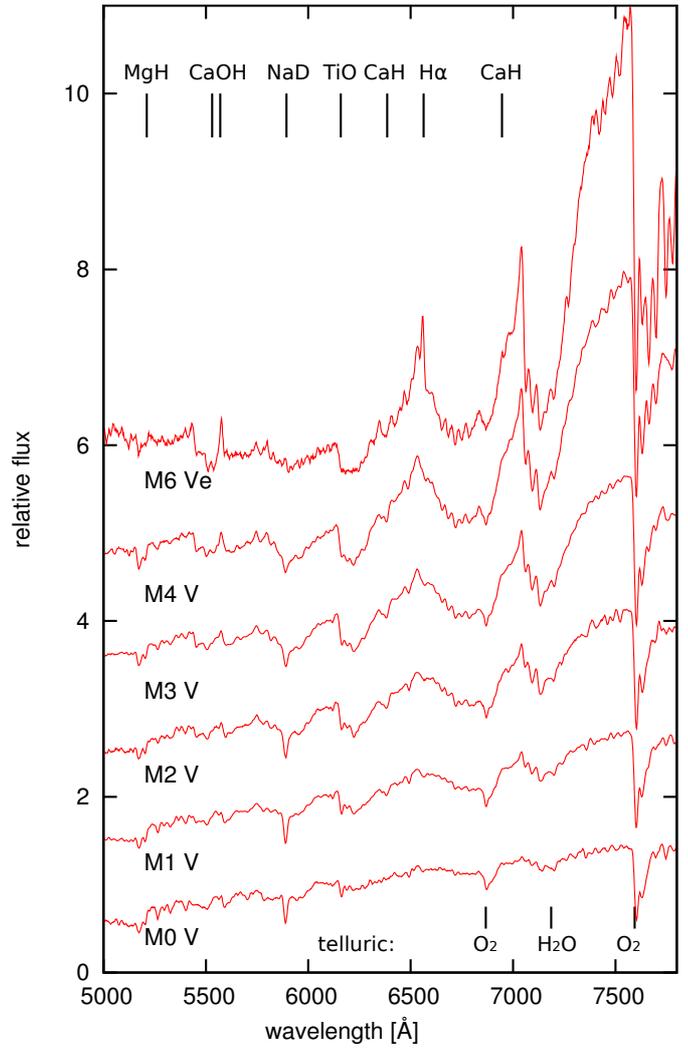}
}
\caption[]{Full sample of average spectra for different spectral types dwarfs according to our classification. The positions of some of the most characteristic features as well as some telluric bands are shown with black lines. Among these are the sodium doublet (not resolved), the MgH, and the CaH bands. The spectra are shifted along the y-axis with an arbitrary constant for presentation purposes.
}
\label{fig:MV_Averages}
\end{figure}

\section{Full spectrophotometric catalogue}

\onecolumn

{\scriptsize 
\begin{longtable}{ccccccccccccc}
\caption{Full spectrophotometric catalogue.}\\
\hline
ID & X & Y  & RA & DEC & r &R& J & H & K & Sp. Cl.& Sp. Cl.&Quality flag \\
   & [px]& [px] & [deg] & [deg] & [deg] &[mag]& [mag] & [mag] & [mag] & photometry & spectroscopy  & \\
\hline
\endfirsthead
\hline
 ID & X & Y  & RA & DEC & r &R& J & H & K & Sp. Cl.& Sp. Cl.&Quality flag \\
   & [px]& [px] & [deg] & [deg] & [deg] &[mag]& [mag] & [mag] & [mag] & photometry & spectroscopy  & \\
\hline
\endhead
 94208&  219.732& 1679.789 &	296.349976& $-$14.828187&   0.1121527&19.772& 17.877& 17.0450& 16.898&   M    &  M1III	   &4	   \\  
 94979&  330.734& 1801.989 &	296.357117& $-$14.834526&   0.1206048&19.509& 17.528& 16.7200& 16.515&   Mca  &  M1III	   &4	   \\
 96968&  608.073& 1462.465 &	296.337067& $-$14.850516&   0.1073543&18.978& 16.924& 16.0800& 15.693&   C    &  C5.5	   & 5     \\
 97590&  696.476& 1751.644 &	296.354126& $-$14.855531&   0.1249199&19.406& 17.334& 16.5210& 16.218&   M    &  C5.5	   & 5     \\
 98478&  830.038&  930.973 &	296.306183& $-$14.863027&  0.08862444&23.062& 17.451& 16.2840& 15.489&   C    &  --	   & 4     \\
 99641& 1001.033& 2013.256 &	296.369537& $-$14.872893&   0.1464667&20.196& 17.025& 15.9710& 15.317&   C    &  C6.5	   & 5     \\
100035& 1068.920&  652.043 &	296.289734& $-$14.876258&  0.08785845&17.078& 17.404& 16.5370& 15.991&   C    &  --	   & 4     \\
101911& 1355.722&  158.286 &	296.260742& $-$14.892557&  0.09137722&20.645& 17.952& 17.1530& 16.985&   M    &  M	   & 4     \\
\hline
\end{longtable}
\par}

\twocolumn

\end{appendix}

\end{document}